\documentclass[secnumarabic,amssymb, nobibnotes, aps, prc]{revtex4} %, showpacs, showkeys

\usepackage{graphicx}
\usepackage{amssymb}
\usepackage{amsmath}
\usepackage{bm}
\usepackage{kotex}
\usepackage{times}
\usepackage{xcolor}
\usepackage{ulem}
\usepackage{cancel}
\usepackage{verbatim}
\usepackage{pifont}
\usepackage{rotating}
\usepackage{subfigure}
\usepackage{multirow}
\usepackage{changepage}
\usepackage{mathtools}

\usepackage[colorlinks=true,linkcolor=red,citecolor=blue]{hyperref}

\newcommand{\Qa}{Q_{\alpha}}
\newcommand{\Qae}{Q_{\alpha}^{\rm exp}}

\newcommand{\Pa}{P_{\alpha}}
\newcommand{\Pae}{P_{\alpha,{\rm exp}}}
\newcommand{\Za}{Z_{\alpha}}

\newcommand{\Zd}{Z_{\rm d}}
\newcommand{\Aa}{A_{\alpha}}

\newcommand{\Ad}{A_{\rm d}}

\newcommand{\be}{\begin{equation}}
\newcommand{\ee}{\end{equation}}
\newcommand{\bea}{\begin{eqnarray}}
\newcommand{\eea}{\end{eqnarray}}

\newcommand{\minus}{\scalebox{0.75}[1.0]{$-$}}

\begin{document}
\setcounter{page}{1}

\title{ $\alpha$-decay half-lives for even-even isotopes of W to U }%by deformed relativistic Hartree-Bogoliubov theory in continuum } within relativistic mean-field models
\author{Yong-Beom \surname{Choi}}
\email{1991.yb.choi@gmail.com}
\affiliation{Center for Innovative Physicist Education and Research, Extreme Physics Institute, and Department of Physics, Pusan National University, Busan 46241, Korea}
\author{Chang-Hwan \surname{Lee}}
\email{clee@pusan.ac.kr}
\affiliation{Department of Physics and Center for Innovative Physicist Education and Research, Pusan National University, Busan 46241, Korea}
\author{Myeong-Hwan \surname{Mun}}
%\email{aa3101@gmail.com}
\affiliation{ Department of Physics and Origin of Matter and Evolution of Galaxy Institute, Soongsil University, Seoul 06978, Korea }
\author{Soonchul \surname{Choi}}
%\email{scchoi0211@ibs.re.kr}
\author{Youngman \surname{Kim}}
%\email{ykim@ibs.re.kr}
\affiliation{ Center for Exotic Nuclear Studies, Institute for Basic Science, Daejeon 34126, Korea }

\date[]{}

\begin{abstract}
We investigate $\alpha$-decay half-lives  for 74 $\le$ Z $\le$ 92 even-even nuclei within the semiclassical WKB approximation in deformed relativistic Hartree-Bogoliubov theory in continuum (DRHBc). The $\alpha$-particle preformation factors are estimated from cluster-formation model using both empirical AME2020 binding energies and numerical ones obtained by a deep neural network (DNN) study in which the available DRHBc binding energies are used as training set.
We find that our estimated $\alpha$-decay half-lives are qualitatively in agree with experimental results.
We also compare our results with the empirical formulae, ZZCW and UNIV.
Based on these observation, we extend our predictions of $\alpha$-decay half-lives for the isotopes whose experimental data are not available.
\end{abstract}

%\pacs{????}
%\keywords{????}

\maketitle

%{\color{red} Version: \today}

%======================================	

\section{Introduction}

Worldwide, many radioactive ion beam (RIB) facilities have been built or are under construction, and 
new opportunities for the deep understanding of the nuclear structure are emerging.
New RIB facilities include the RIB Factory (RIBF) at RIKEN in Japan~\cite{Kamigaito:2020pll}, the Facility for Rare Isotope Beams (FRIB) in the United States of America~\cite{Davide2022}, the High Intensity heavy-ion Accelerator Facility (HIAF) in China~\cite{Zhou:2022pxl}, the Facility for Antiproton and Ion Research (FAIR) in Germany~\cite{Sturm:2010yit}, and the Rare isotope Accelerator complex for ON-line experiments (RAON) in Korea~\cite{Jeong:2018qvi}.
%These facilities will enable us to study the shell structure of atomic nuclei and to explore the limits of their existence,
With the help of these new RIB facilities, more exotic nuclei and their structure can be investigated, and the limits of the existence of exotic nuclei can be explored.

Alpha($\alpha$)-decay has been used as one of the important tools to understand the nuclear structure information.
%
%In order to study them effectively, it is necessary to theoretically study the radioactive decays. 
The $\alpha$-decay %, one of the radioactive decays, 
can be relatively easily studied by calculating the half-lives of nuclei.
A method of calculating the $\alpha$-decay half-lives using the nuclear density distribution constructed is to adopt the semiclassical WKB approximation. In the WKB approximation, the potentials between nucleons are used in the calculation of the half-lives. %This method enables us to avoid the direct incorporation of the nuclear density. 
As an example, cosh nuclear potential~\cite{Buck:1992zz} is employed to calculate the $\alpha$-decay half-lives~\cite{Deng:2017ids, Sun:2016bbw}. 
To consider the neutron and proton density distributions within the WKB approximation framework, the density-dependent cluster model with two-parameter Fermi distribution
%, called deformed Woods-Saxon distribution, 
has been adopted in various studies~\cite{Xu:2005hlv, Xu:2005ukj, Ismail:2010zza, Ni:2010zza, Coban:2012zz, Maroufi:2019txx, Rojas-Gamboa:2022zva, Wang:2022axn}.

%
%In addition, corrections such as surface diffuseness have been applied because two-parameter Fermi distribution 
%cannot describes the neutron and proton density distributions near the surface properly~\cite{Maroufi:2019txx, Wang:2022axn}.

In this work, deformed relativistic Hartree-Bogoliubov theory in continuum (DRHBc) is used to calculate the density distributions.
The DRHBc was developed to predict the properties of exotic and stable nuclei by self-consistently incorporating the pairing correlation and continuum effects~\cite{Zhou:2009sp, Li:2012gv, Meng:2015hta, Zhang:2020wvp, DRHBcMassTable:2022uhi, DRHBcMassTable:2022rvn}.
The pairing interaction is essential to describe open-shell model. Thus, Hartree-Fock with the pairing interaction, such as BCS, had been employed to understand $\alpha$-decay near shell closures as well as away from shell closure. However, since Hartree-Fock with BCS is not enough to properly handle exotic nuclei near or away from shell closure~\cite{Teran:2002ax}, the relativistic Hartree-Bogoliubov framework has been developed to provide a better description of the exotic nuclei far from $\beta$-stability~\cite{Vretenar:2005zz}. 
In exotic nuclei, neutron or proton Fermi energy is close to continuum threshold. continuum effects are also treated in the DRHBc.
In order to expand the applicability of the DRHBc, angular momentum projection~\cite{Sun:2021nfb, Sun:2021nyl}, finite amplitude method~\cite{Sun:2022gdu}, and two-dimensional collective Hamiltonian~\cite{Sun:2022qck} have been implemented recently.
The DRHBc with PK1~\cite{Long:2003dn} or PC-PK1 density functional~\cite{Zhao:2010hi} is widely used to explore various attractive exotic phenomena, such as deformed halo structure~\cite{Sun:2018ekv, Sun:2020tas}, neutron drip line shift by incorporating axial deformation degrees of freedom~\cite{In:2020asf}, bound nuclei beyond two-neutron drip line~\cite{Zhang:2021ize, Pan:2021oyq, He:2021thz}, bubble structure~\cite{Choi:2022rdj}, and shape coexistence~\cite{Choi:2022rdj, Kim:2021skf}.
In this study, we employ the DRHBc with PC-PK1 density functional to predict neutron or proton densities.

To calculate the $\alpha$-decay half-lives, the preformation factor $P_\alpha$, the probability of the $\alpha$-particle formation in the parent nucleus, should be also considered.
Since it is still difficult to calculate microscopically the preformation factor in relativistic density functional theory, the cluster-formation model was suggested to estimate $\Pa$~\cite{SalehAhmed:2013nwa, Ahmed:2015kra}. 
In the cluster-formation model, $\Pa$ is relatively easy to estimate because only the binding energies of parent nucleus and its neighboring nuclei are required. 
In this model, $\alpha$-particle formation inside the parent nucleus is considered to be the quantum mechanical cluster formation. 
The $P_\alpha$ of $^{212}$Po in the cluster-formation model is estimated to be 0.221~\cite{Ahmed:2015kra}. 
Considering that the $P_\alpha$ of $^{212}$Po in the microscopic calculation is 0.3~\cite{Varga:1992zz}, the preformation factor estimated by this model is acceptable.
In the systematic studies using cluster-formation model, the unpaired nucleons in odd-$A$ or odd-odd nuclei are expected to be the origin of the reduction in $P_\alpha$ due to the Pauli blocking~\cite{Deng:2015qha, Deng:2016ibo, Wan:2021wny}. 
It is also found that $P_\alpha$ in magic number nuclei decreases. %~\cite{Deng:2015qha, Deng:2016ibo}.
We estimate $P_\alpha$ using the cluster-formation model~\cite{SalehAhmed:2013nwa}.
Since the investigation of mass table for the odd-$Z$ nuclei in the DRHBc is still in progress, $\Pa$ of even-even nuclei cannot be estimated systematically at the moment.
To obtain $\Pa$ of even-even nuclei, we use the binding energies obtained through the results of deep neural network (DNN) study~\cite{CK2}.

We calculate the $\alpha$-decay half-lives for even-even isotopes of W to U using the predicted densities in the DRHBc within the WKB approximation framework. 
In this study only $\alpha$-decay half-lives of even-even nuclei are considered, and therefore, the hindrance by the Pauli blocking will not be studied~\cite{Deng:2015qha, Deng:2016ibo, Wan:2021wny}.
This article is organized as follows. In Sec.~\ref{Tf}, the neutron and proton density distributions by the DRHBc is briefly introduced, and
the framework of the WKB approximation for calculating $\alpha$-decay half-lives is discussed. 
In Sec.~\ref{HLresults}, we compare the preformation factors obtained using the AME2020 data and deep-learning results, and show the estimated $\alpha$-decay half-lives.
We discuss the relation between including axial deformation degrees of freedom and the difference between neutron Fermi energy and proton Fermi energy.
We finally summarize and discuss this work in Sec.~\ref{HLs&d}.

%----------------------------------------------------

\section{Theoretical framework} \label{Tf}

\subsection{Deformed relativistic Hartree-Bogoliubov theory in continuum }\label{densitiy_distribution}
In this section, we briefly introduce how the DRHBc includes deformations and pairing correlation. 
The $\alpha$-decay half-lives in the WKB approximation require density distributions for a daughter nucleus and the $\alpha$-particle. 
We introduce how the density distributions are described in the DRHBc.

The relativistic mean-field theory is built from the Lagrangian density~\cite{Ring:1996qi}.
The Lagrangian density for point-coupling is described in detail in Ref.~\cite{A_textbook_of_Jie_Meng}.
The relativistic Hartree-Bogoliubov equation which incorporates the mean-field and pairing field simultaneously is obtained with the variational method and the Bogoliubov transformation~\cite{Kucharek:1991arbi},
\begin{gather}
     \begin{pmatrix}
          h_{\rm D} - \lambda_\tau & \Delta \\ - \Delta^* & -h^*_{\rm D} + \lambda_\tau
     \end{pmatrix}
     \begin{pmatrix}
         U_k \\ V_k
     \end{pmatrix}
     = E_k \,
     \begin{pmatrix}
         U_k \\ V_k
     \end{pmatrix}, \label{HF_eq}
\end{gather}
where $h_{\rm D}$ is Dirac Hamiltonian for the nucleons, $\lambda_\tau$ represents the Fermi energy of a nucleon, $\Delta$ is the pairing field, $U_k$ and $V_k$ correspond to the quasiparticle wave functions, and $E_k$ means the energy of a quasiparticle state $k$. 
The pairing potential for particle-particle channel is used
\bea \label{pairing_potential1}
\Delta_{k k^{'}}(\bm{r},\bm{r^{'}}) = - \sum_{\tilde{k}\tilde{k}^{'}} \,
                                 V^{pp}_{k k^{'},\tilde{k}\tilde{k}^{'}} (\bm{r},\bm{r^{'}})
                                 \kappa_{\tilde{k}\tilde{k}^{'}} (\bm{r},\bm{r^{'}}),
\eea
where, $k$, $k^{'}$, $\tilde{k}$, and $\tilde{k}^{'}$ denote the quasiparticle states and the pairing tensor is defined by $\kappa=V^{\ast}U^T$~\cite{A_textbook_of_Peter_Ring}. 
For pairing interaction in the particle-particle channel $V^{pp}$, in the DRHBc, the density-dependent zero-range pairing interaction is used
\bea
V^{pp} (\bm{r},\bm{r^{'}}) = \frac{V_0}{2} \left(1 - P^{\sigma} \right)
                             \delta (\bm{r} - \bm{r^{'}} )
                             \left(1 - \frac{\rho(\bm{r})}{\rho_{\rm sat}} \right),
\eea
where $\rho_{\rm sat}$ is the nuclear saturation density. 
The Dirac Hamiltonian $h_{\rm D}$ for the nucleons is given as~\cite{A_textbook_of_Jie_Meng}
\be
    h_{\rm D} = \bm{\alpha} \cdot \bm{p} \, + \, \beta \left(M+S(\bm{r})\right) \, + \, V(\bm{r})
\ee
where $S(\bm{r})$ and $V(\bm{r})$ are scalar and vector potentials, which are expressed as 
\bea
   S(\bm{r}) &=& \alpha_{\rm S} \rho_{\rm S} + \beta_{\rm S} \rho^2_{\rm S} +
               \gamma_{\rm S} \rho^3_{\rm S} + \delta_{\rm S} \Delta\rho_{\rm S}, \label{sPot}  \nonumber \\
   V(\bm{r}) &=& \alpha_V \rho_V + \gamma_V \rho^3_V + \delta_V \Delta \rho_V
             + e A_0  +  \alpha_{\rm TV} \tau_3 \rho_{\rm TV} + \delta_{\rm TV} \tau_3 \Delta \rho_{\rm TV}  . \label{vPot}
\eea
The local densities $\rho_{\rm S}(\bm{r})$, $\rho_{\rm V}(\bm{r})$, and $\rho_{\rm TV}(\bm{r})$ are represented as
\bea
\rho_{\rm S}(\bm{r}) = \sum_{k>0} \, \bar{V_k}(\bm{r}) V_k(\bm{r}), \nonumber \\
\rho_{\rm V}(\bm{r}) = \sum_{k>0} \, V^\dag_k (\bm{r}) V_k(\bm{r}),\nonumber  \\
\rho_{\rm TV}(\bm{r}) = \sum_{k>0} \, V^\dag_k (\bm{r}) \tau_3  V_k(\bm{r}),
\eea
where the summation for $k$ is performed only in positive energy in Fermi sea with the no-sea approximation. 
In the DRHBc, spatial reflection symmetry is preserved. Thus, the densities ($\rho_{\rm S}(\bm{r})$, $\rho_{\rm V}(\bm{r})$, $\rho_{\rm TV}(\bm{r})$) are expanded using Legendre polynomials as~\cite{Price:1987sf}
\be
\rho(\bm{r}) = \sum_{\lambda_{\rm L}} \, \rho_{\lambda_{\rm L}} (r) P_{\lambda_{\rm L}} (\cos \theta).
            \,\,\, \lambda_{\rm L} = 0, \, 2, \, 4, \, \cdots.
\ee

For the DRHBc, as described in Ref.~\cite{Choi:2022rdj}, the pairing strength $V_0=-325$ MeV$\cdot {\rm fm}^3$ and the angular momentum cutoff for the Dirac Woods-Saxon basis $J_{\rm max}=23/2~\hbar$ are taken. 
Other are described in Ref.~\cite{DRHBcMassTable:2022uhi}.

\subsection{$\alpha$-decay in WKB approximation\label{WKB_equations}}

\begin{figure}[t]
\begin{center}
\includegraphics[width=0.45\columnwidth]{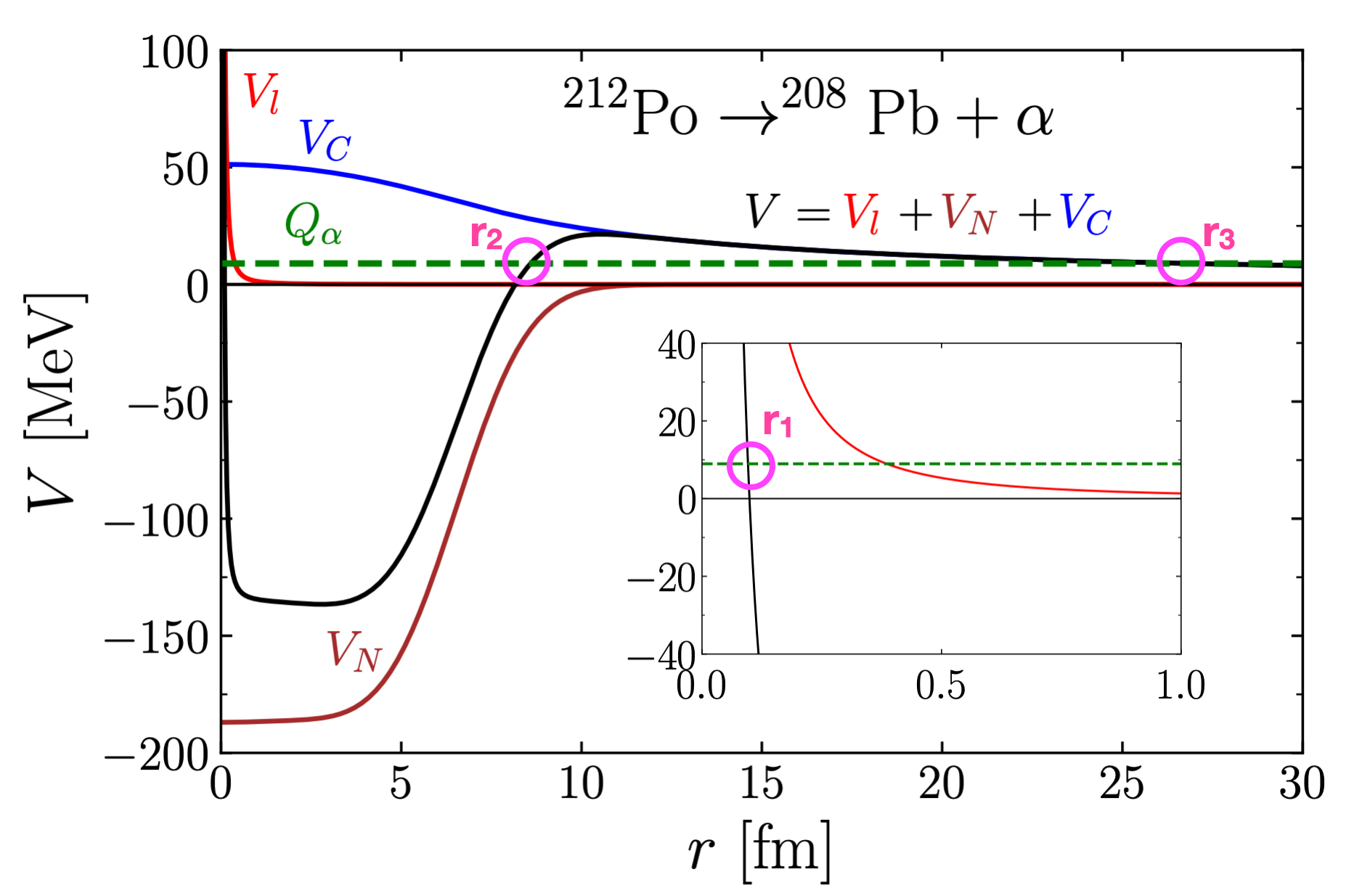}
\caption{The potentials between $^{208}$Pb and $\alpha$-particle as a function of $r$. The angular momentum $l=0$ is adopted. The double-folding potential $V_N$ is effective in the range of $r\le 10$ fm. If a daughter nucleus is axially defomed, the potential between the daughter nucleus and $\alpha$-particle is considered as a function of $r$ and the angle from the symmetry axis $\beta$.} \label{R2_and_R3_example} 
\end{center}
\end{figure}

\begin{figure}[t]
\begin{center}
\includegraphics[width=0.4\columnwidth]{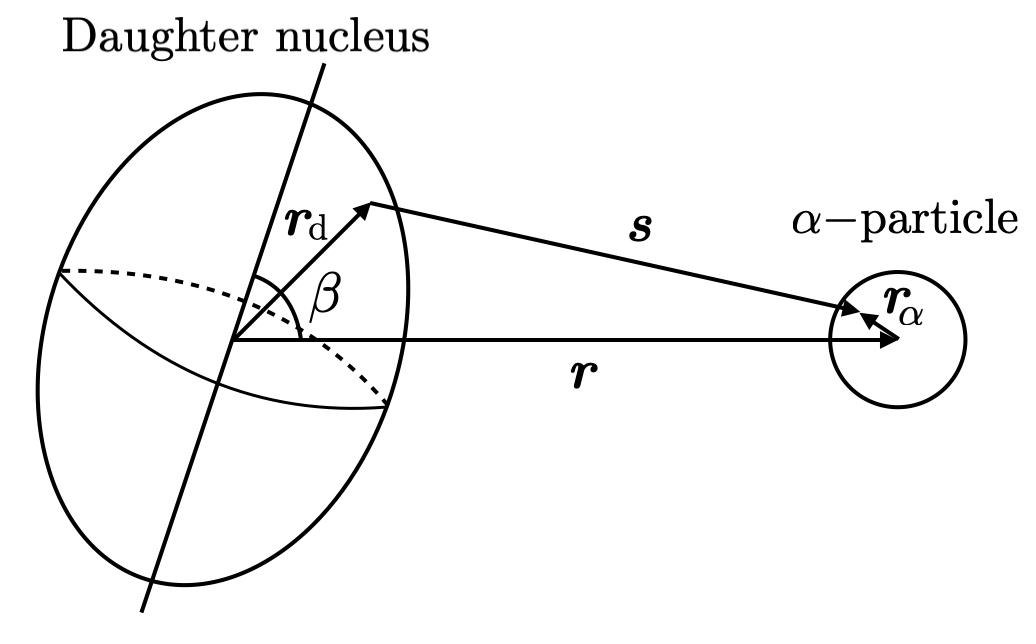}
\caption{Schematic picture to describe the coordinates used in the double-folding model. The letter $\beta$ is for the orientation angle between the $\alpha$-particle and the symmetry axis of the daughter nucleus, and ${\bm s} = {\bm r} - {\bm r}_{\rm d} + {\bm r}_\alpha$. } \label{Schematic_plot}
\end{center}
\end{figure}

The $\alpha$-decay process is understood as the formation of $\alpha$-particle inside the parent nucleus and penetrating the potential barrier of the daughter nucleus
with a given $Q$-value $\Qa$ as in Fig.~\ref{R2_and_R3_example}. 
Therefore, the $\alpha$-decay half-life can be calculated within the WKB approximation as~\cite{Gurvitz:1986uv, Xu:2006cr}
\bea
T_{1/2} &=& \frac{\hbar \ln 2}{\Gamma} \label{T_halp_with_WKB} \label{hl_eq}, \\
\Gamma &=& P_\alpha N_{\rm f} \frac{\hbar^2}{4\mu} P_{\rm total} , \label{eq_gamma}
\eea
where $\Gamma$, $\Pa$, $N_{\rm f}$, and $P_{\rm total}$ are decay width, preformation factor of $\alpha$-particle, normalization factor for bound-stated wave function, and the total penetration probability, respectively. 
To calculate the decay width $\Gamma$, the preformation factor $P_\alpha$ must be first calculated.
%in addition to other factors in Eq.~(\ref{eq_gamma}), in the framework of DRHBc, one can discuss the reliability of the nuclear distributions calculated by DRHBc by comparing calculated $\alpha$-decay half-lives with experimental data. 
However, the preformation factor $P_\alpha$ in Eq.~(\ref{eq_gamma}) cannot yet be calculated microscopically in the DRHBc framework.
Instead we calculate $P_\alpha$ as in the cluster-formation model \cite{SalehAhmed:2013nwa}.
The preformation factor of $\alpha$-particle $\Pa$ for even-even nuclei is calculated in the cluster-formation model~\cite{Ahmed:2015kra, Deng:2015qha} as
\be
P_\alpha = \frac{2S_{\rm p} + 2S_{\rm n} - S_\alpha}{S_\alpha},
\label{eq_preformation}
\ee
where $S_{\rm p}$, $S_{\rm n}$, $S_\alpha$ are one-proton, one-neutron, and $\alpha$-particle separation energies, respectively.
Separation energies can be obtained as
\bea
S_{\rm p}(Z,N) &=& E_{\rm b}(Z,N)-E_{\rm b}(Z-1,N), \\
S_{\rm n}(Z,N) &=& E_{\rm b}(Z,N)-E_{\rm b}(Z,N-1), \\
S_\alpha(Z,N) &=& E_{\rm b}(Z,N)-E_{\rm b}(Z-2,N-2),
\eea
where $E_{\rm b}(Z,N)$ is the binding energy of given proton number $Z$ and neutron number $N$. 
We take $E_{\rm b}(Z,N)$ from both AME2020 and the results of the DNN study~\cite{CK2} since, at
the moment, the binding energies for odd-$A$ nuclei are not available in the DRHBc framework.
The normalization factor and the total penetration probability for axially deformed nuclei in Eq.~(\ref{eq_gamma}) are given as
\bea
N_{\rm f} &=& \frac{1}{2} \int_{0}^{\pi} N_{\rm f}(\beta) \sin\beta d\beta,\\
P_{\rm total} &=& \frac{1}{2} \int_{0}^{\pi} \exp \bigg[ -2 \int_{r_2(\beta)}^{r_3(\beta)} k(r',\beta) d{r}' \bigg] \sin \beta d\beta, \label{Gamma_halp_with_WKB}
\eea
where the normalization factor $N_{\rm f}(\beta)$ for the orientation angle between the $\alpha$-particle and the symmetry axis of the daughter nucleus $\beta$ and $k(r,\beta)$ are expressed as
\bea
N_{\rm f}(\beta) &=& \bigg[ \int_{r_1(\beta)}^{r_2(\beta)} \frac{d{r}'}{k(r',\beta)} \cos^2 \bigg( \int_{r_1(\beta)}^{r'} d{r}'' k(r'',\beta) - \frac{\pi}{4} \bigg) \bigg]^{-1} 
\simeq \bigg[ \int_{r_1(\beta)}^{r_2(\beta)} \frac{d r'}{2k(r',\beta)} \bigg]^{-1}, \\ 
k(r,\beta) &=& \sqrt{\frac{2\mu}{\hbar^2} |Q_\alpha-V(r,\beta)| }, 
\eea
and $\mu$ is the reduced mass of the $\alpha$-particle and daughter nucleus,  $r_{1,2,3}(\beta)$ are the classical turning points which are marked in Fig.~\ref{R2_and_R3_example},  and $\beta$ is the azimuthal angle between the symmetry axis of daughter nucleus and the direction of the produced $\alpha$-particle as in Fig.~\ref{Schematic_plot}. Note that the outer turning point $r_3$ is determined mostly by the Coulomb potential. We take empirical $\Qae$ from AME2020~\cite{Wang:2021xhn}.

%----- Fig. 1 ---------------
%-------------------------------

%----- Fig. 2 ---------------
%-------------------------------

In the density-dependent cluster model, the potential $V(r,\beta)$ consists of centrifugal potential $V_l(r)$, nucleus-nucleus double-folding potential $V_{\rm N}(r,\beta)$, and Coulomb potential $V_{\rm C}(r,\beta)$~\cite{Wan:2021wny},
as plotted in Fig.~\ref{R2_and_R3_example},
\be
V(r,\beta) = V_l(r) + V_{\rm N}(r,\beta) + V_{\rm C}(r,\beta).
\ee
The centrifugal potential is expressed using the Langer modified form~\cite{Langer:1937qr, Buck:1992zza, Buck:1993sku}
\be
V_l(r)= \frac{\hbar^2}{2\mu} \frac{(l+1/2)^2}{r^2},
\ee
where $l$ is the quantum number corresponding to the orbital angular momentum transferred by $\alpha$-particle. 
As we only consider the $\alpha$-decay of even-even nuclei, we apply $l=0$~\cite{Xu:2006fq}. 
The nuclear and Coulomb potentials are built from double-folding model \cite{Bertsch:1977sg, Satchler:1979ni, Kobos:1982pdw, Kobos:1984zz}:
\bea
V_{\rm N}(r,\beta) &=& \lambda \int d{\bm r}_{\rm d} d{\bm r}_\alpha \rho_{\rm d} ({\bm r}_{\rm d})  \rho_\alpha ({\bm r}_\alpha) v({\bm s}), \label{NN_potential} \\
V_{\rm C}(r,\beta) &=& \int d{\bm r}_{\rm d} d{\bm r}_\alpha \rho_{\rm d}^{\rm proton} ({\bm r}_{\rm d})  \rho_\alpha^{\rm proton} ({\bm r}_\alpha) \frac{e^2}{s}, \label{charge_potential} 
\eea
where we take $e^2 = 1.4399764$ MeV$\cdot$fm and plot ${\bm s} = {\bm r} + {\bm r}_\alpha - {\bm r}_{\rm d}$ in Fig.~\ref{Schematic_plot}.
We set the density distribution of $\alpha$-particle as standard Gaussian form $\rho_\alpha (r) = 0.4229 \exp(-0.7024r^2)$ \cite{Satchler:1979ni}.
From Eq.~(\ref{Gamma_halp_with_WKB}), it can be seen that $\alpha$-decay half-lives depend largely on the wave number $k(r,\beta)$ in the range of $r =r_2 \sim r_3$. 
We use the M3Y effective nucleon-nucleon interaction $v({\bf s})$ in MeV unit,
\be
v({\bm s}) = 7999e^{-4s}/(4s) - 2134e^{-2.5s}/(2.5s) - 276(1-0.005Q_\alpha/A_{\alpha}) \delta({\bm s}),
\ee
for considering the double-folding potential~\cite{Bertsch:1977sg, Kobos:1982pdw}. In the above equation, $A_{\alpha}$ is the mass number of $\alpha$-particle and $\delta$ is the delta function.

To calculate $V_{\rm C}$ and $V_{\rm N}$, we use $\rho_{\rm d}$ and $\rho_{\rm d}^{\rm proton}$ calculated from the DRHBc;
\bea
V_{\rm N\ \textrm{or}\ C}(r , \beta) &=& \sum_{\lambda_{\rm L}=0,2,4,...} V_{{\rm N\ \textrm{or}\ C}, \lambda_{\rm L}}(r , \beta).
\eea
By using the Fourier transformation, one can derive $V_{\rm N,\lambda_L}$ and $V_{\rm C,\lambda_L}$ as
\bea
V_{{\rm N}, \lambda_{\rm L}}(r , \beta) &=& \frac{2}{\pi} \int_{0}^{\infty} dk k^2 j_{\lambda_{\rm L}}(kr) \tilde{\rho}_\alpha(k) \bigg[ \int^{\infty}_{0} dr_{\rm d} r_{\rm d}^2 \rho_{{\rm d}, \lambda_{\rm L}}(r_{\rm d}) j_{\lambda_{\rm L}} (k r_{\rm d}) \bigg] \tilde{v}(k) P_{\lambda_{\rm L}} (\cos \beta), \\
V_{{\rm C}, \lambda_{\rm L}}(r , \beta) &=& 8e^2 \int_{0}^{\infty} dk j_{\lambda_{\rm L}}(kr) \tilde{\rho}_\alpha^{\rm proton}(k) \bigg[ \int^{\infty}_{0} dr_{\rm d} r_{\rm d}^2 \rho_{{\rm d}, \lambda_{\rm L}}^{\rm proton}(r_{\rm d}) j_{\lambda_{\rm L}} (k r_{\rm d}) \bigg] P_{\lambda_{\rm L}} (\cos \beta),
\eea
where $j_{\lambda_{\rm L}}(kr)$ is the spherical Bessel function. 
The terms $\tilde{\rho}_\alpha(k), \tilde{\rho}_{{\rm d}, \lambda_{\rm L}}(k)$, and $\tilde{v}(k)$ are Fourier transformation of $\rho_\alpha(r), \rho_{{\rm d}, \lambda_{\rm L}}(r)$, and $v(r)$, respectively. 
The forms of $\tilde{\rho}_\alpha(k)$ and $\tilde{v}(k)$ are written in Appendix A of Ref.~\cite{Bai:2018hbe}.
The generalized equation is written in Ref.~\cite{1983ZPhyA.310..287R}.
The property $\exp(i {\bm k} \cdot {\bm r}) = \sum_{\lambda_{\rm L}} i^{\lambda_{\rm L}} (2\lambda_{\rm L}+1) j_{\lambda_{\rm L}} (kr) P_{\lambda_{\rm L}} (\cos \theta)$ is used for applying Fourier transformation. 
We take $\lambda_{\rm L, max}=8$ to be consistent with the current DRHBc results.

In Eq.~(\ref{NN_potential}), $\lambda$ is the normalization factor and related to the strength of nuclear potential. It is determined by the Bohr-Sommerfeld quantization condition~\cite{Xu:2006fq}%\cite{Buck:1992zza, Buck:1994zz}
\be
\int_{0}^{\pi} \int_{r_1(\beta)}^{r_2(\beta)} \sqrt{ \frac{2\mu}{\hbar^2} [\Qa-V(r,\beta)] } \sin \beta dr d\beta  = (2n+1)\frac{\pi}{2} = (G-L+1) \frac{\pi}{2}, \label{BS_qc}
\ee
where $r_1(\beta),~r_2(\beta)$ are the set of classical turning points ($\Qa=V(r,\beta)$), and $\Qa$ and $G$ are the $\alpha$-decay energy and global quantum number, respectively. 
Note that there is no additional constraint to fix the global quantum number $G$. 
We take $G$ as follows~\cite{Buck:1992zz, Buck:1996zza}
\bea
&&G = 22 ~ (N_{\rm  par} > 126), \nonumber \\
&&G = 20 ~ (82 < N_{\rm par} \le 126), \nonumber
%&&G = 18 ~ (N_{\rm par}>126 \le 82), \nonumber \\
\eea
where $N_{\rm par}$ is the neutron number of parent nucleus.

\section{Results\label{HLresults}}

\subsection{Preformation factor}

%----- Fig. 3 --------------
\begin{figure}[t]
\begin{center}
\subfigure[]{\includegraphics[width=0.45\columnwidth]{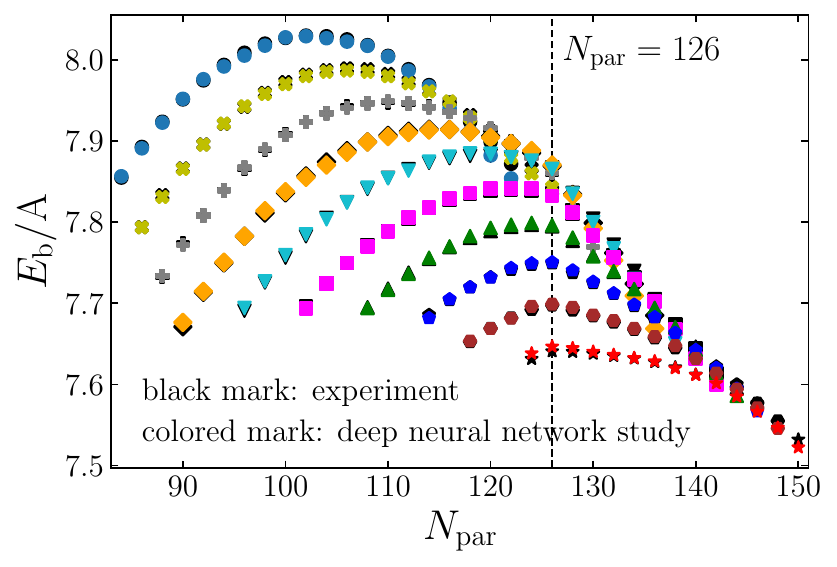}}
\subfigure[]{\includegraphics[width=0.45\columnwidth]{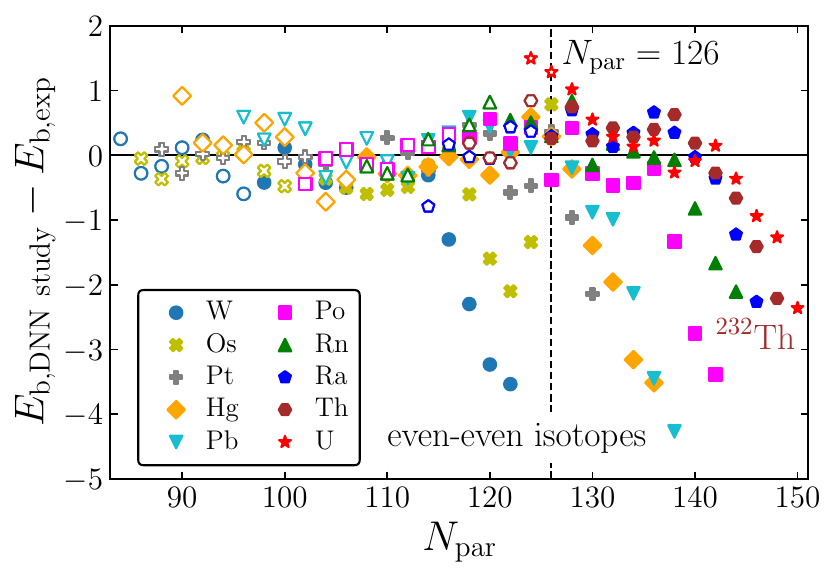}}
\subfigure[]{\includegraphics[width=0.45\columnwidth]{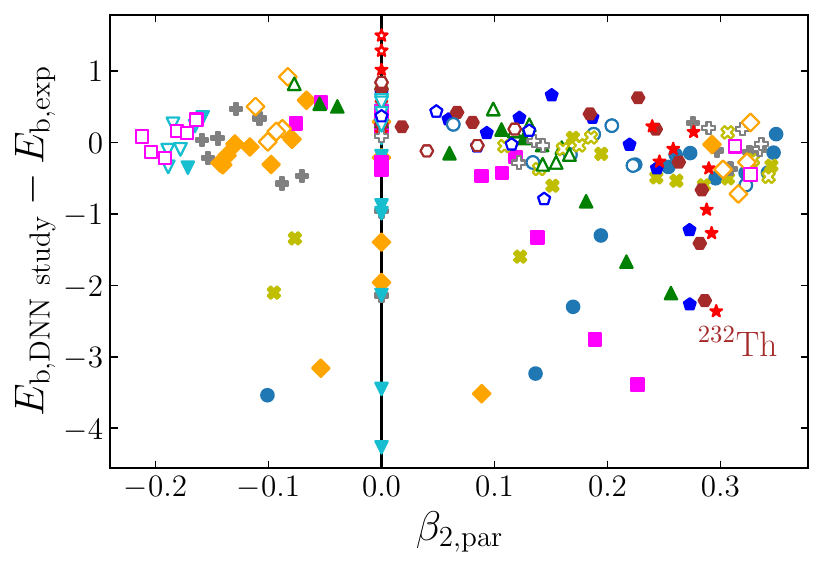}}
\subfigure[]{\includegraphics[width=0.45\columnwidth]{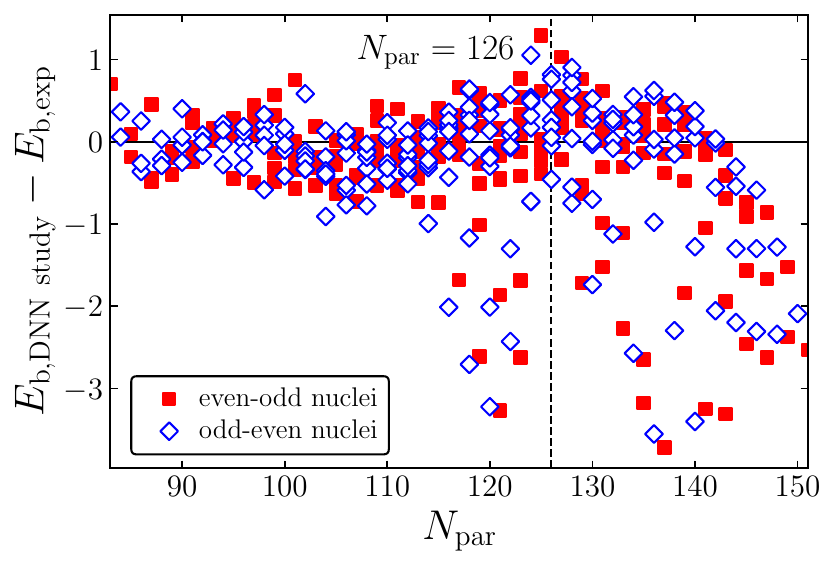}}
\caption{Binding energies for selected nuclei. (a): Comparison of binding energies per nucleon between AME2020 and deep neural network (DNN) study in which the DRHBc results are used as training set  as a function of parent neutron number $N_{\rm par}$ for even-even isotopes. (b): Difference in the binding energies between DNN study and AME2020 for even-even nuclei. (c): Difference in the binding energies as a function of the quadrupole deformation $\beta_{2,\rm par}$ of the parent nucleus. 
(d): same as (b) for even-odd and odd-even isotopes.
Nuclei corresponding to each color in (a-c) are summarized in (b). 
The empty marks in (b) and (c) represent nuclei with Fermi energy difference larger than 6 MeV as shown in Fig.~\ref{fig_DRHBc_lambda}. Deformation parameters $\beta_{2,{\rm par}}$ in the DRHBc are summarized in Fig.~\ref{fig_DRHBc_beta}.
}  \label{DNNS_BE_figure}
\end{center}
\end{figure}
%-------------------------------

In order to obtain the decay width, in Eq.~(\ref{eq_gamma}), the preformation factor $P_\alpha$ has to be provided.
As discussed in the previous section, we calculate $P_\alpha$ using binding energies by following the cluster-formation model \cite{SalehAhmed:2013nwa} as defined in Eq.~(\ref{eq_preformation}).
Since, at the moment, the binding energies for odd-even and odd-odd nuclei are not available in the DRHBc framework, $S_{\rm n}$ and $S_{\rm p}$ cannot be obtained in the DRHBc. 
Therefore, we use the binding energies obtained from a DNN study~\cite{CK2} in which available results from the DRHBc calculations and the AME2020 data are used as a training set.  
When the binding energies are available both in the DRHBc and AME2020, those of AME2020 were used in the training set.
To obtain the odd-even and odd-odd binding energies using a deep neural network, the binding energies of even-even and even-odd isotopes from the DRHBc calculations are used as a training set. In addition, to include some information about odd-odd and odd-even isotopes to the deep neural network the authors of ~\cite{CK2} also use the binding energies in AME2020~\cite{Wang:2021xhn} as a training set. The inputs of the neutral network are the proton number ($Z$), neutron number ($N$), nuclear pairing ($\delta$) and shell effect ($P$) which are defined by
%$\delta$ is defined by 
\begin{eqnarray}
\delta &=&\frac{(-1)^Z+(-1)^N}{2}, \\ % \;\;\;\;\;  %, and $P$ is given by
%\begin{equation}
P &=& \frac{\nu_P\nu_n}{\nu_P+\nu_n}\, ,
\end{eqnarray}
where $\nu_P$ ($\nu_N$) is the difference between the proton (neutron) number of an isotope of interest and the nearest magic numbers.
For more details of the neutral network to predict the odd-odd and odd-even masses based on the DRHBc and AME2020, we refer to~\cite{CK2}.
For the comparison, we also use the experimental binding energy and $\Qae$ from AME2020~\cite{Wang:2021xhn}.
Binding energies used in this work are summarized in Tables~\ref{hl_tbl} and \ref{hl_tbl2}.

Figure~\ref{DNNS_BE_figure}~(a) summarizes the binding energy per nucleon for the selected even-even nuclei obtained by the DNN study. 
For the comparison, experimental binding energies~\cite{Wang:2021xhn} are plotted with black color.
In Fig.~\ref{DNNS_BE_figure}~(b), differences in the binding energies between the DNN study and experimental results are given as a function of the parent neutron number. 
For nuclei with small neutron number ($N_{\rm par}< 126$), the differences between the DNN study and experiment  are small compared to nuclei with large neutron number ($N_{\rm par} \ge 126$). This may indicate that DNN study results are more consistent with experiment values for the nuclei with $N_{\rm par} <$ 126.
In Fig.~\ref{DNNS_BE_figure}~(c), the differences in the binding energies between the DNN study and experimental results are summarized as a function of quadrupole deformation $\beta_{2,\rm par}$. This figure shows no strong correlation between the binding energy of the DNN study and the deformation.
%Nuclei with spherical shape ($\beta_{2,\rm par}=0$) show most wide spread.
In Fig.~\ref{DNNS_BE_figure}~(d), the differences for even-odd and odd-even nuclei are summarized. 
Overall, it shows similar behavior as even-even nuclei in Fig.~\ref{DNNS_BE_figure}~(b), but the deviations of the DNN study results from experimental values for even-odd nuclei are relatively larger than those for odd-even nuclei. 
It may indicate that the DNN study is more sensitive to neutron distributions than proton distributions.

%----- Fig. 4 -----------------
\begin{figure}[t]
\begin{center}
\includegraphics[width=0.45\columnwidth]{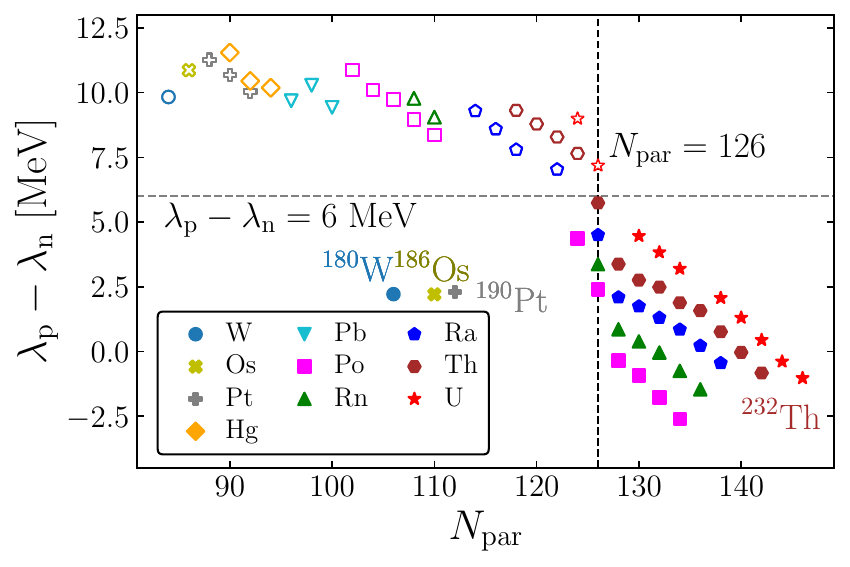}
\caption{Fermi energy difference between protons and neutrons in the DRHBc. Vertical dashed line correspond to the neutron magic number $N_{\rm par}=126$. Horizontal dashed line corresponds to $\lambda_p-\lambda_n = 6$ MeV. 
Empty marks are used for nuclei above the horizontal dashed line. 
}
\label{fig_DRHBc_lambda}
\end{center}
\end{figure}
%-------------------------------

%----- Fig. 5 -----
\begin{figure}[t]
\begin{center}
%\subfigure[]{\includegraphics[width=0.45\columnwidth]{Figures/fig5/beta2_vs_Np.pdf}}
%\subfigure[]{\includegraphics[width=0.45\columnwidth]{Figures/fig5/beta2_vs_Np_2.pdf}}
\includegraphics[width=0.45\columnwidth]{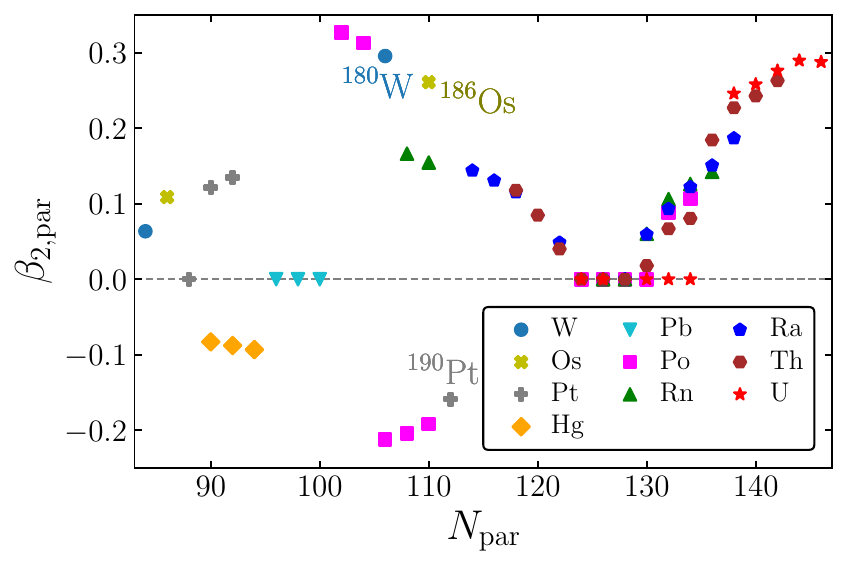}
\caption{ 
Quadrupole deformation of parent nuclei $\beta_{2,{\rm par}}$ in the DRHBc. }
\label{fig_DRHBc_beta}
\end{center}
\end{figure}
%-------------------------------

%In order to determine whether there is a correlations between the preformation factors and nuclear properties in DRHBc, the Fermi energy differences and deformation parameters in DRHBc are summarized in Figs.~\ref{fig_DRHBc_lambda} and \ref{fig_DRHBc_beta}.  
Figure 4 shows the Fermi energy differences and deformation parameters in the DRHBc.
The Fermi energy differences are larger than 6 MeV for most of nuclei with $N_{\rm par} < 126$ except for $^{180}$W, $^{186}$Os, $^{190}$Pt, and $^{208}$Po as in Fig.~\ref{fig_DRHBc_lambda}. 
The experimental $\alpha$-decay half-lives of $^{180}$W, $^{186}$Os, and $^{190}$Pt are important because the half-lives of nearby nuclei are dominated by other decay modes, such as $\beta$-decay.
%For nuclei with less Fermi energy difference ($\lambda_{\rm p}$ - $\lambda_{\rm n}$) and $N_{\rm par} < 126$, $\alpha$-decay may occurs, but other decays, including $\beta$-decay, occurs or dominates.
%Although there are only three nuclei ($^{180}$W, $^{186}$Os, and $^{190}$Pt) whose $\alpha$-decay half-lives can be compared with the calculation, which is important in this study.
The $\alpha$-decay half-lives of $^{180}$W, $^{186}$Os, and $^{190}$Pt are longest in this study ($10^{25.7}$, $10^{22.8}$, and $10^{19.183}$ s, respectively).
These nuclei can be used for verifying whether the half-life calculation provides consistent results for long experimental $\alpha$-decay half-lives.
Also, these nuclei can be used to verify whether the half-life for defomed nuclei is accurate since quadrupole deformations of these nuclei are $\beta_{2,\rm par}=0.243$ for $^{180}$W, $\beta_{2,\rm par}=0.209$ for $^{186}$Os, and $\beta_{2,\rm par}=0.163$ for $^{190}$Pt, respectively~\cite{Moller:2015fba}.
The relation, $|\beta_{2,\rm par}|$ for $^{180}$W $>$ $|\beta_{2,\rm par}|$ for $^{186}$Os $>$ $|\beta_{2,\rm par}|$ for $^{190}$Pt, can also be confirmed in Fig.~\ref{fig_DRHBc_beta}.
In this figure, one can confirm $\beta_{2,{\rm par}} = 0.0$ for the nuclei with neutron magic number $N_{\rm par}=126$. 
%The results in Fig.~\ref{fig_DRHBc_beta} indicate that the deformations are very important.
%However, in Fig.~\ref{DNNS_BE_figure} (c), there seems to no direct correlation between the binding energy differences and the deformation as discussed before.

%----- Fig. 6 ---------------
\begin{figure}[t]
\begin{center}
\subfigure[]{\includegraphics[width=0.45\columnwidth]{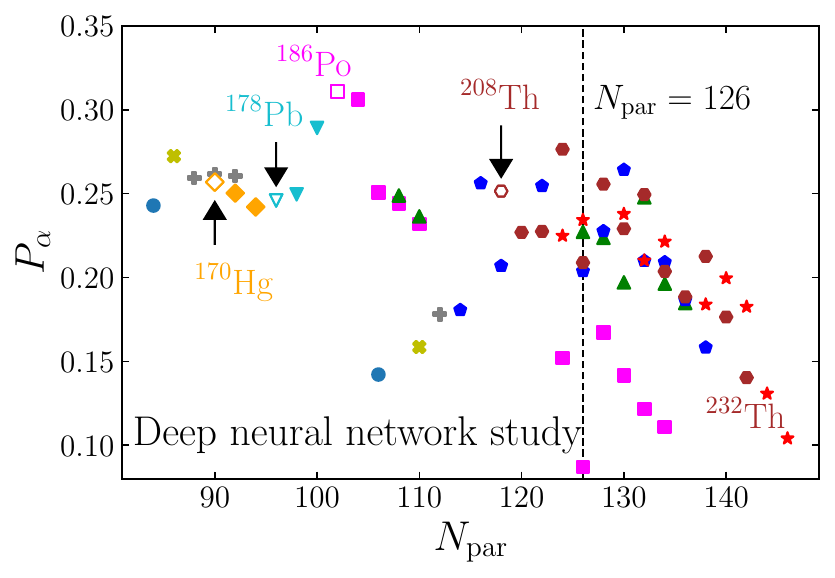}}
\subfigure[]{\includegraphics[width=0.45\columnwidth]{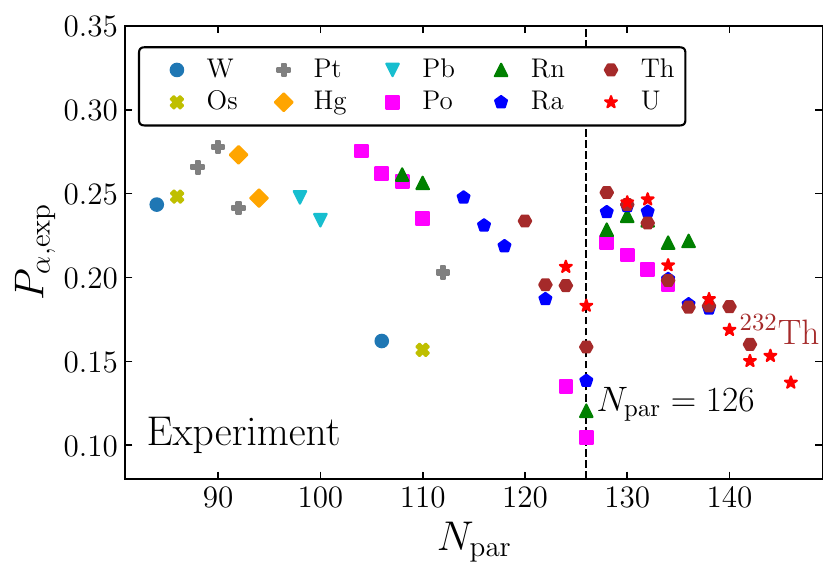}}
\subfigure[]{\includegraphics[width=0.45\columnwidth]{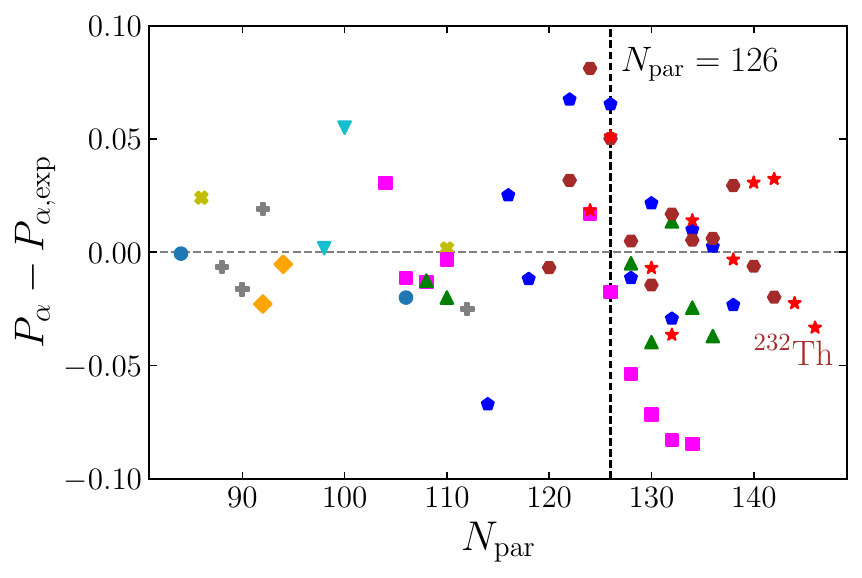}}
\caption{Preformation factors as a function of parent neutron number $N_{\rm par}$.
(a) Estimated $\Pa$ from DNN study. (b) Estimated $\Pa$ with experimental binding energies (AME2020).
In (a) nuclei with empty mark in ($^{170}$Hg, $^{178}$Pb, $^{186}$Po, $^{208}$Th) correspond to the nuclei for which some of the experimental results are not available to obtain $P_\alpha$. (c) The difference between $\Pa$ of (a) and (b). 
}
\label{Pa_figure}
\end{center}
\end{figure}
%-------------------------------

In Fig.~\ref{Pa_figure} (a) and (b), preformation factors obtained with the DNN study~\cite{CK2} and experimental data (AME2020) are summarized. In Fig.~\ref{Pa_figure} (a), nuclei with empty marks ($^{170}$Hg, $^{178}$Pb, $^{186}$Po, $^{208}$Th) correspond to the nuclei for which experimental results are not enough to obtain $P_\alpha$. Note that binding energies for $^{169}$Hg, $^{177}$Pb, $^{185}$Po, and $^{207}$Th are missing in AME2020.
In Fig.~\ref{Pa_figure} (b), $P_\alpha$ are relatively small at the magic number $N_{\rm par}=126$ and  increase
as neutron number increases to $N_{\rm par}=128$, independently of nuclei; i.e., $P_{\alpha,{\rm exp}}(N_{\rm p}=128) > P_{\alpha,\rm exp}(N_{\rm p}=126)$. Results with AME2012 also show the similar behavior~\cite{Deng:2015qha}. Except Rn, such a trend exists for the result of the DNN study as  in Fig.~\ref{Pa_figure} (a).
The differences between the DNN study and experimental results are summarized in Fig.~\ref{Pa_figure}~(c).
In this figure, below the neutron magic number of parent nuclei $N_{\rm par}=126$, $^{202,210}$Ra isotopes show the largest $P_\alpha$ diffenrece in the isotopes.
For the magic number $N_{\rm par}=126$, $^{210}$Po shows smallest preformation factor in the DNN study and experiment.
Above the magic number, $^{214, 216, 218}$Po give largest differences in $P_\alpha$.
If one compare Fig.~\ref{Pa_figure} with Figs.~\ref{fig_DRHBc_lambda} and \ref{fig_DRHBc_beta}, 
there are no direct correlations between $P_\alpha$ and the properties of nuclei, such as the Fermi energy difference and the deformation parameter $\beta_{2,{\rm par}}$.

%--------------------

\subsection{$\alpha$-decay half-lives}
In this section, we present our calculated $\alpha$-decay half-lives with given preformation factors as discussed in the previous section.
In order to assess the quality of our results, we use experimental binding energy and $\Qae$ from AME2020~\cite{Wang:2021xhn}.
For the comparison with experiment, we take the experimental $\alpha$-decay half-lives from NUASE2020~\cite{Kondev:2021lzi}. 
Because several decay modes can exist for each nucleus, we consider only 67 experimental data 
for which the branching ratio of $\alpha$-decay are greater than 99\%.
The calculated $\alpha$-decay half-lives are also summarized in Tables~\ref{hl_tbl} and \ref{hl_tbl2}.

%----- Fig. 7 -----
\begin{figure}[t]
\begin{center}
\subfigure[]{\includegraphics[width=0.45\columnwidth]{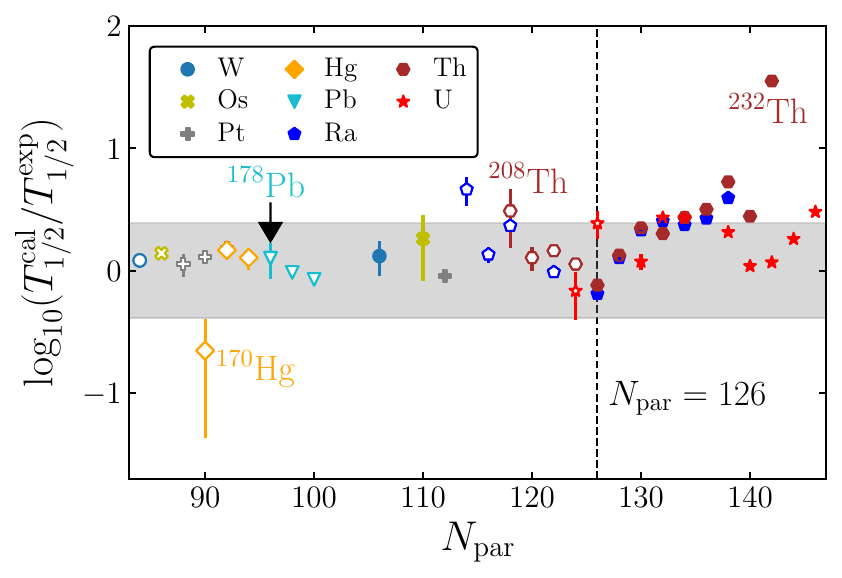}}
\subfigure[]{\includegraphics[width=0.45\columnwidth]{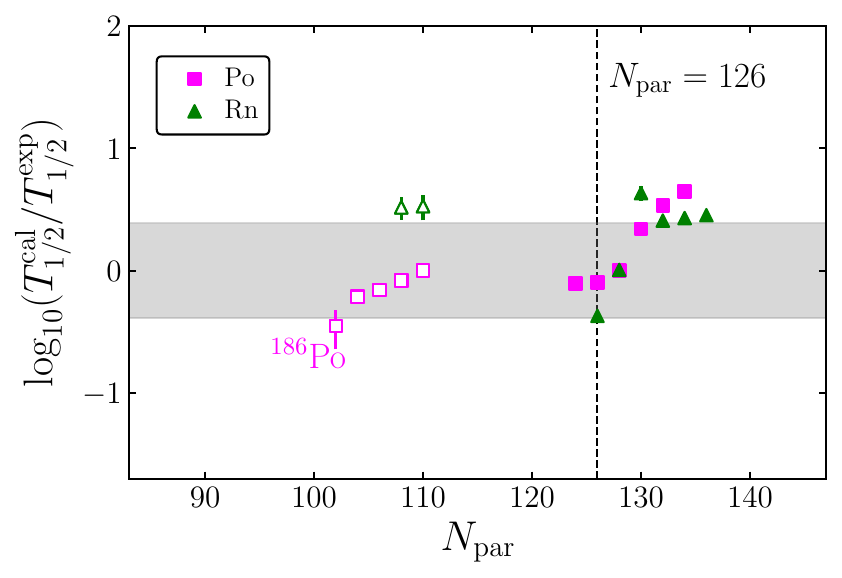}}
%\subfigure[]{\includegraphics[width=0.45\columnwidth]{Figures//fig4/Hl_comparision_x_N_par.pdf}}
\caption{Our calculated $\alpha$-decay half-lives as a function of parent neutron number $N_{\rm par}$.
For the easy comparison, Po and Rn are shown in (b) separately.
Note that proton-rich nuclei (empty marks) correspond to $\lambda_p-\lambda_n > 6$ MeV as in Fig.~\ref{fig_DRHBc_lambda} and stable nuclei (filled marks) correspond to $\lambda_p-\lambda_n < 6$ MeV as in Fig.~\ref{fig_DRHBc_lambda}.
Vertical dashed lines correspond to the neutron magic number $N_{\rm par}=126$. %
The width of the horizontal shaded region represents $2\sigma$ where $\sigma = 0.437$ is the standard deviation defined in Eq.~(\ref{eq_sigma}).
} \label{fig_T_N}
\end{center}
\end{figure}
%-------------------------------

%-- Fig.8-----------------
\begin{figure}[t]
\begin{center}
\subfigure[]{\includegraphics[width=0.45\columnwidth]{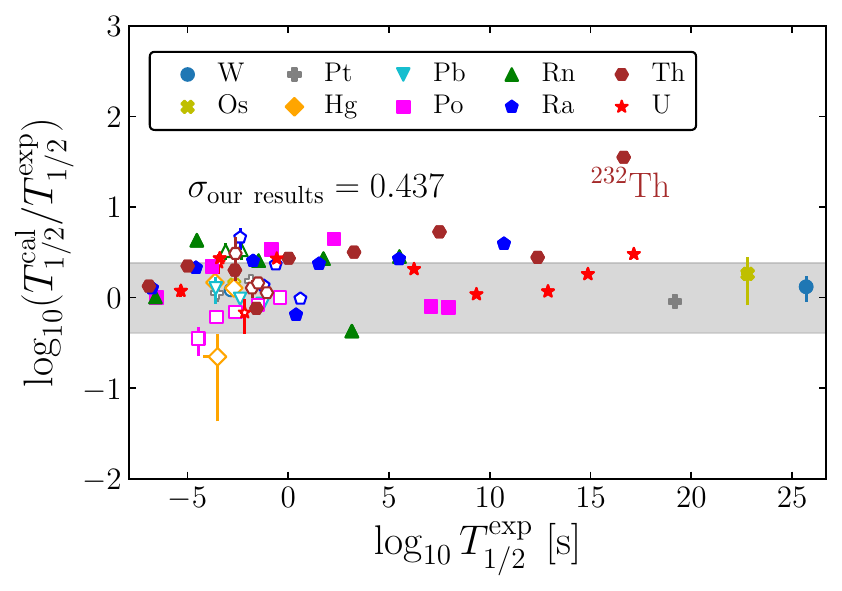}}
\subfigure[]{\includegraphics[width=0.45\columnwidth]{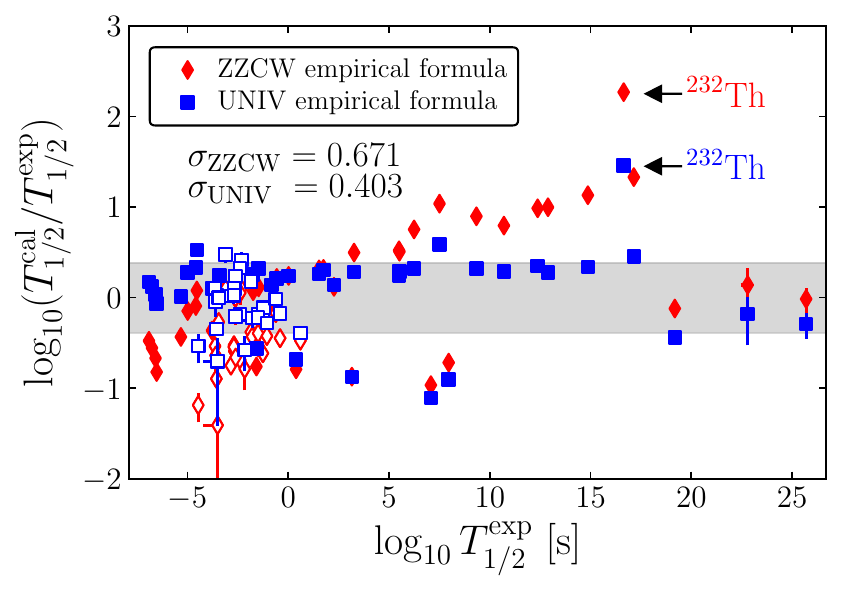}}
\caption{Comparison of our calculated $\alpha$-decay half-lives and those with empirical formulae (ZZCW and UNIV) with experimental results. 
(a) comparison between the results of the DRHBc combined with preformation factors estimated by the DNN study results and experimental values. (b) comparison between the results of empirical formulae and experimental results.
Horizontal shaded regions are the same as in Fig.~\ref{fig_T_N} with standard deviation $\sigma=0.437$.
The calculated half-lives for $^{232}$Th show the biggest discrepancies with the experimental data even in calculations using empirical formulae.
}  \label{comparison_with_ef}
\end{center} 
\end{figure}
%--------------------------

%---- Fig.9 ------------------
\begin{figure}[t]
\begin{center}
\subfigure[]{\includegraphics[width=0.45\columnwidth]{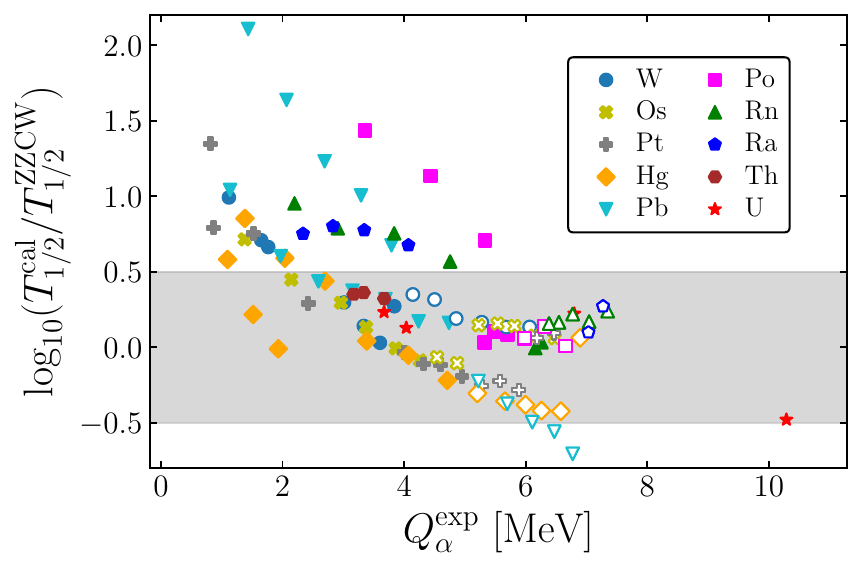}}
\subfigure[]{\includegraphics[width=0.45\columnwidth]{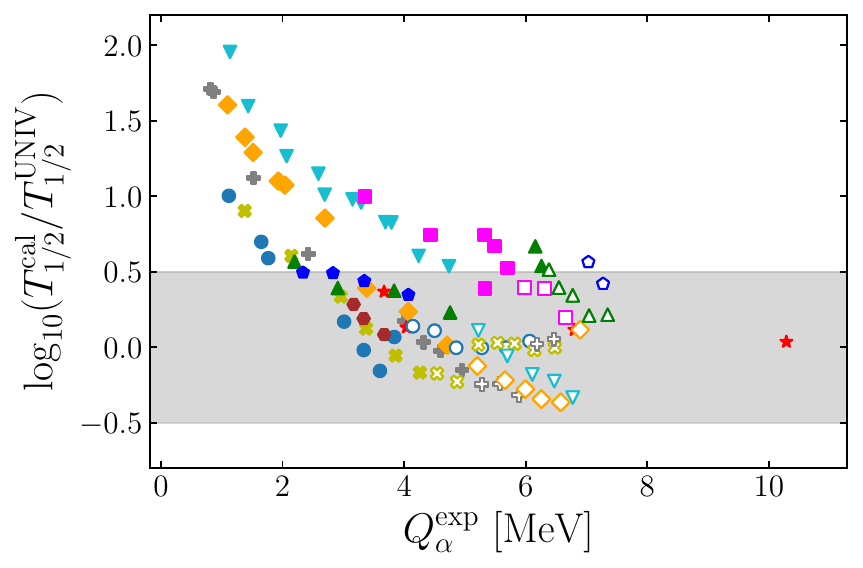}}
\subfigure[]{\includegraphics[width=0.45\columnwidth]{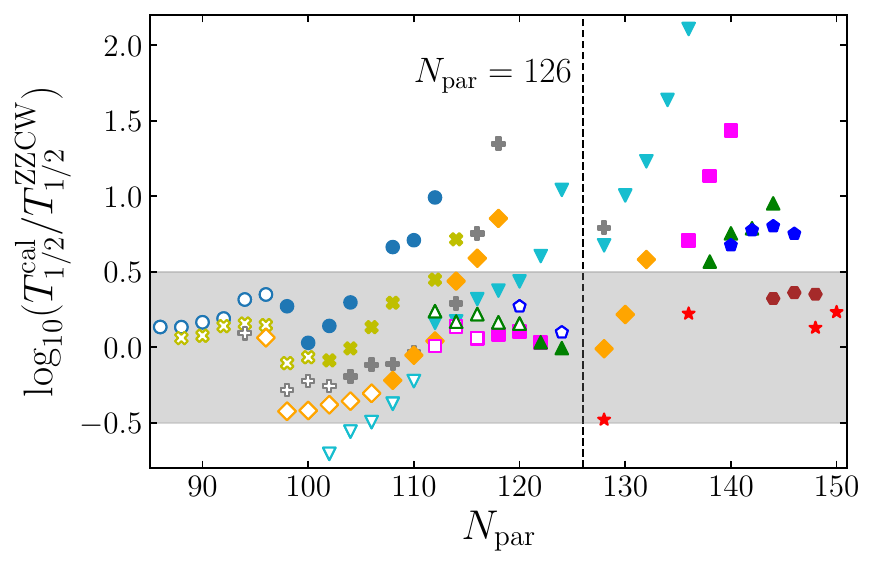}}
\subfigure[]{\includegraphics[width=0.45\columnwidth]{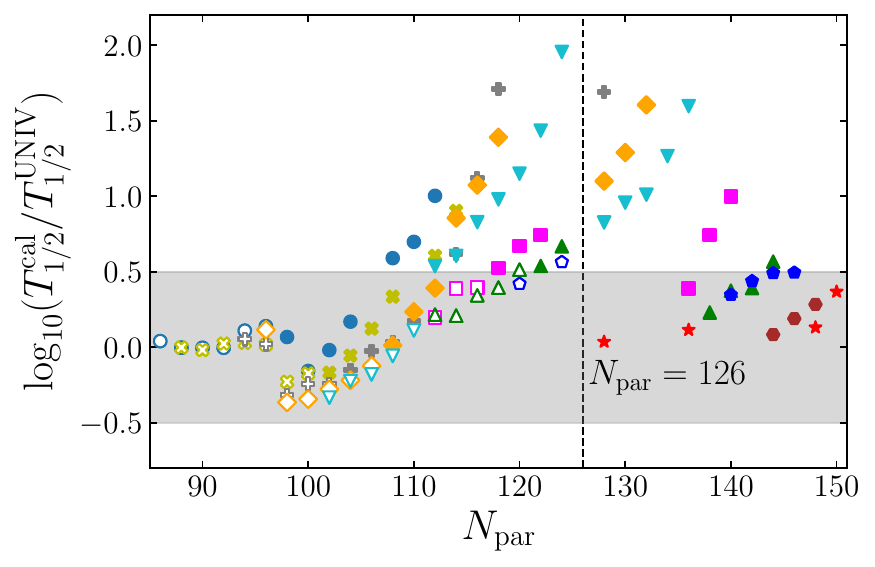}}
\caption{Comparison of our calculated $\alpha$-decay half-lives between our approach and the empirical formulae (ZZCW and UNIV) for the nuclei whose $\alpha$-decay half-lives are not measured experimentally. (a) as a function of $Q_\alpha$ and (b) as a function of parent neutron number $N_{\rm par}$. Empty marks represent proton-rich nuclei.
In this plot, all isotopes with $Q_\alpha>0$ between W and U are included.
} \label{prediction_and_comparison_with_ef}
\end{center}
\end{figure}
%-----------------

%------- Fig. 7 discussion --------------

In Fig.~\ref{fig_T_N}, our calculated $\alpha$-decay half-lives are compared with experimental values.
In order to compare our calculated $\alpha$-decay half-lives with experimental data, we evaluate the standard deviation $\sigma$,
\be
\sigma=\sqrt{\frac{1}{N} \sum_{i=1}^{N}\left( \log_{10}T_{1/2,i}^{\rm cal} - \log_{10}T_{1/2,i}^{\rm exp}\right)^2},
\label{eq_sigma}
\ee
which are plotted as horizontal shaded regions in Fig.~\ref{fig_T_N}.
Overall, considering the uncertainties in the binding energy estimations by the DNN study, our calculated $\alpha$-decay half lives are consistent with the experimental values within $\pm\sigma$. %except for a few cases.
If we compare Fig.~\ref{fig_T_N} with Fig.~\ref{fig_DRHBc_beta}, the calculated $\alpha$-decay half lives are consistent with experimental values independently of deformation parameter $\beta_{2,{\rm par}}$.
Below the neutron magic number $N_{\rm par}=126$, $^{170}$Hg, $^{186}$Po, $^{202}$Ra, and $^{208}$Th lie out of the shaded region.
The reason for this stems partly from limitation in $\Pa$ calculations.
Note that $P_{\alpha, {\rm exp}}$ for $^{170}$Hg, $^{186}$Po, and $^{208}$Th are not available because binding energies for $^{169}$Hg, $^{185}$Po, and $^{207}$Th are not available in AME2020.
In the DRHBc, $\lambda_{\rm p}>0$ for $^{170}$Hg, $^{186}$Po, and $^{208}$Th, it may be more difficult to measure the binding energies of $^{169}$Hg, $^{185}$Po, and $^{208}$Th.
The predictions from theoretical models such as the DRHBc will be important to study such nuclei.
The discrepancy in the $\alpha$-decay half-life of $^{202}$Ra is partly because $\Pa < \Pae$. 
Incorrectly estimated $\Pa$ is not the only cause of discrepancies in the half-lives.
For example, the difference in preformation factor of $^{210}$Ra is large, but the half-life comparison result is very accurate.
Beyond the neutron magic number of $130<N_{\rm par}$ for Ra and Th isotopes, the predicted $\alpha$-decay half-lives exceed experimental values, as shown in Fig.~\ref{fig_T_N} (a). 
This mismatch can be hardly explained by overestimated $\Pa$ and further study needs to be conducted.
For $130<N_{\rm par}<136$ of Po and Rn isotopes (see Fig.~\ref{fig_T_N} (b)), our calculated $\alpha$-decay half-lives are larger than experimental values. 
This is partly because our estimated $\Pa$ is smaller than the experimental value as in Fig.~\ref{Pa_figure}.

%------- Fig. 8 discussion --------------

In Fig. \ref{comparison_with_ef}, the comparion of our calculated $\alpha$-decay half-lives with the experimental $\alpha$-decay results (AME2020) is shown as function of the experimental $\alpha$-decay results.
In addition, the comparison of the results with empirical formulae, ZZCW~\cite{Zhang:2017pwm} and UNIV~\cite{Poenaru:2011zz} with the experimental $\alpha$-decay results is also presented as function of the experimental $\alpha$-decay results.
The details of the empirical formulae are summarized in the Appendix and results from our study with DNN and empirical formulae are summarized in Tables~\ref{hl_tbl_pd} and \ref{hl_tbl_pd2}.
In Fig. \ref{comparison_with_ef} (b) $\sigma_{\rm ZZCW}$ = 0.671 and $\sigma_{\rm UNIV}$ = 0.403 which are comparable to $\sigma=0.437$ for our DRHBc$+$DNN study calculation. 
If we consider only proton-rich nuclei (empty marks in Fig.~\ref{comparison_with_ef}), $\sigma_{\rm DRHBc+DNN study}=0.405$ and $\sigma_{\rm UNIV}$ = 0.467. This may indicate that DRHBc$+$DNN study is reasonably consistent with experiment for both stable and proton-rich nuclei.
$^{232}$Th, the most stable isotope of Th, is exceptional because the estimated $\alpha$-decay half-lives by empirical formulae as well as our method are much larger than experiment.
Further study is required to investigate this largest discrepancy.
Based on these observations, we extend our study to the nuclei whose $\alpha$-decay half-lives are not measured experimentally. 

%------- Fig. 9 discussion --------------

In Fig.~\ref{prediction_and_comparison_with_ef}, we compare our predicted $\alpha$-decay half-lives with empirical formulae results for the nuclei whose $\alpha$-decay half-lives are not available in experiments.
The differences between our prediction and empirical formulae become significant as $Q_\alpha$ decreases.
The difference is significant in the range roughly $110<N_{\rm par}<140$ mainly due to small $\Qa$.
Note that our calcultions are consistent with empirical formulae results within $\pm 0.7$ for Ra, Th, and U isotopes.
For Pb isotopes, the differences between our results and the empirical formulae tends to be more apparent.

\section{Summary \label{HLs&d}}

In this work, we calculated $\alpha$-decay half-lives of 74 $\le$ Z $\le$ 92 even-even isotopes in the framework of the semiclassical WKB approximation with the DRHBc.

In order to calculate the $\alpha$-decay half-lives, it is first necessary to estimate $\Pa$. 
%It is still challenge to microscopically estimate $\Pa$. 
By adopting the cluster-formation model, we estimated $\Pa$ using the binding energies of the parent nucleus and its neighboring nuclei.
Since the mass table for odd-$Z$ nuclei in the DRHBc is in progress, we used the binding energies obtained by DNN study.
In this study, the available DRHBc binding energies for even-$Z$ nuclei, and the AME2020 are used as training set for the missing odd-$Z$ nuclei binding energies.
When the binding energies are available both in the DRHBc and AME2020, those of AME2020 were used in the training set.
Comparing the binding energies obtained by DNN study with AME2020 binding energies for both even-$Z$ and odd-$Z$ nuclei, we found that the binding energies obtained by DNN study are consistent with the AME2020 data within $\sim 0.1 \%$ uncertainty.
Using the binding energies from the DNN study, we calculated $P_\alpha$.

We found that our estimated $\alpha$-decay half-lives are qualitatively in agreement with experimental data (NUBASE2020).
In our calculation, since there exist several decay modes for each nucleus, we took only 67 experimental data for which the branching ratios of $\alpha$-decay modes are greater than 99$\%$.
%To qunatatibly compare the experimental data, we adopt the standard deviation $\sigma$. %standard deviations
The standard deviation $\sigma$ between our results and experimental data is 0.437, and
those between empirical formulae and experimental data are 0.671 for ZZCW and 0.403 for UNIV.
The $\alpha$-decay half-life prediction for $^{232}$Th, the most stable isotope of Th, shows largest discrepancy both in our approach and in the empirical formulae.
We noticed that the differences in $\alpha$-decay half-lives estimated in our approach and experimental data
 are not so sensitive to quadrupole deformation of the parent nucleus $\beta_{2,{\rm par}}$.

Based on these observation, we extended our predictions of $\alpha$-decay half-lives for the isotopes whose experimental data are not available. For these isotopes, we compare our results with those obtained using the empirical formulae ZZCW and UNIV, and found that the standard deviations of half-lives for the isotopes with small $\Qa$ increase roughly by a factor of two.

In this study, we focused on $\alpha$-decay half-lives by using the density distributions calculated from the DRHBc within the WKB approximation. Extension to one-proton emission within the same approximation is in progress.

%=======	Acknowledgements ===========================	

\section*{ACKNOWLEDGMENTS}
The authors are grateful to the members of the DRHBc Mass Table Collaboration for useful discussions, to Kaiyuan Zhang for providing the DRHBc even-even mass table data, and to Seonghyun Kim for providing some of the DRHBc results.
YBC and CHL were supported by National Research Foundation of Korea (NRF) grant funded by the
Korea government (Ministry of Science and ICT and Ministry of Education)  (No. 2023R1A2C1005398).
M.-H.M. was supported by the NRF grant funded by the Korea government(MSIT) (Grants No. NRF-2020R1A2C3006177, No. NRF-2021R1F1A1060066, and No. NRF-2021R1A6A1A03043957).
Y. K. was supported in part by the Institute for Basic Science (IBS-R031-D1).
This work was supported by the National Supercomputing Center with supercomputing resources including technical support (KSC-2022-CRE-0333, KSC-2023-CRE-0170).

\bibliographystyle{aip}

\pagebreak

\begin{table*}[b]
%\begin{center}
 \caption{The calculated half-lives with our method, ZZCW, and UNIV. The experimental data $\Qae$ and $T^{\rm exp}_{1/2}$ is taken from AME2020~\cite{Wang:2021xhn} and NUBASE2020~\cite{Kondev:2021lzi}, respectively. The units of $\Qae$, $T_{1/2}$, and $E_{\rm b}$ are MeV, s, and MeV$/c^2$, respectively.
\label{hl_tbl} }
 \begin{adjustwidth}{-0.7cm}{}
  \begin{tabular}{c|c|l|c|c|c|c|c|l|l|l}
   \hline
    \hline
 \multirow{2}{*}{\parbox{1cm}{Parent \\nucleus}} & \multicolumn{2}{c|}{Experimental data} & \multirow{2}{*}{\parbox{1.2cm}{$\beta_{2,{\rm p}}$ \\ (DRHBc)}} & \multicolumn{2}{c|}{ $E_{\rm b}$ } & \multicolumn{2}{c|}{Preformation factor} & \multicolumn{3}{c}{ $T^{\rm cal}_{1/2}$ }  \\  
 \cline{2-3}
 \cline{5-11}
  & \phantom{a } $\Qae$ \phantom{a } & \multicolumn{1}{c|}{$T^{\rm exp}_{1/2}$} & & AME2020 & DNN study & AME2020 & DNN study & \multicolumn{1}{c|}{Our result} & \multicolumn{1}{c|}{ZZCW} & \multicolumn{1}{c}{UNIV} \\
    \hline
$^{158}$W &  6.612 & (1.43 $\pm ~ $0.18) $\times ~ 10^{-3}$ & \phantom{$\minus$}0.064\phantom{$\minus$} & 1241.090 & 1241.344  &  0.243  &  0.243 & 1.74 $\times ~ 10^{-3}$ & 2.53 $\times ~ 10^{-4}$ & 1.49 $\times ~ 10^{-3}$  \\
$^{180}$W &  2.515 & (5.02 $\pm ~ $2) $\times ~ 10^{25}$ & \phantom{$\minus$}0.296\phantom{$\minus$} & 1444.580 & 1444.079  &  0.162  &  0.142 & 6.61 $\times ~ 10^{25}$ & 4.80 $\times ~ 10^{25}$ & 2.54 $\times ~ 10^{25}$  \\
$^{162}$Os &  6.768 & (2.1 $\pm ~ $0.1) $\times ~ 10^{-3}$ & \phantom{$\minus$}0.109\phantom{$\minus$} & 1262.628 & 1262.576  &  0.248  &  0.272 & 2.91 $\times ~ 10^{-3}$ & 5.39 $\times ~ 10^{-4}$ & 2.65 $\times ~ 10^{-3}$  \\
$^{186}$Os &  2.821 & (6.3 $\pm ~ $3.5) $\times ~ 10^{22}$ & \phantom{$\minus$}0.261\phantom{$\minus$} & 1484.807 & 1484.272  &  0.157  &  0.159 & 1.15 $\times ~ 10^{23}$ & 8.64 $\times ~ 10^{22}$ & 4.15 $\times ~ 10^{22}$  \\
$^{166}$Pt &  7.292 & (2.94 $\pm ~ $0.62) $\times ~ 10^{-4}$ & \phantom{$\minus$}0.000\phantom{$\minus$} & 1283.678 & 1283.769  &  0.266  &  0.259 & 3.32 $\times ~ 10^{-4}$ & 6.18 $\times ~ 10^{-5}$ & 3.13 $\times ~ 10^{-4}$  \\
$^{168}$Pt &  6.990 & (2.02 $\pm ~ $0.1) $\times ~ 10^{-3}$ & \phantom{$\minus$}0.121\phantom{$\minus$} & 1305.968 & 1305.688  &  0.278  &  0.262 & 2.61 $\times ~ 10^{-3}$ & 6.03 $\times ~ 10^{-4}$ & 2.56 $\times ~ 10^{-3}$  \\
$^{170}$Pt &  6.707 & (1.393 $\pm ~ $0.016) $\times ~ 10^{-2}$ & \phantom{$\minus$}0.135\phantom{$\minus$} & 1327.400 & 1327.420  &  0.241  &  0.260 & 2.12 $\times ~ 10^{-2}$ & 5.79 $\times ~ 10^{-3}$ & 2.09 $\times ~ 10^{-2}$  \\
$^{190}$Pt &  3.269 & (1.52 $\pm ~ $0.01) $\times ~ 10^{19}$ & $\minus$0.159\phantom{$\minus$} & 1509.834 & 1509.868  &  0.203  &  0.178 & 1.38 $\times ~ 10^{19}$ & 1.17 $\times ~ 10^{19}$ & 5.60 $\times ~ 10^{18}$  \\
$^{170}$Hg &  7.773 & (3.1 $\pm ~ $2.5) $\times ~ 10^{-4}$ & $\minus$0.083\phantom{$\minus$} & 1304.070 & 1304.988  & ---  &  0.257  & 6.89 $\times ~ 10^{-5}$ & 1.21 $\times ~ 10^{-5}$ & 6.15 $\times ~ 10^{-5}$  \\
$^{172}$Hg &  7.524 & (2.31 $\pm ~ $0.09) $\times ~ 10^{-4}$ & $\minus$0.088\phantom{$\minus$} & 1326.740 & 1326.931  &  0.273  &  0.250 & 3.40 $\times ~ 10^{-4}$ & 6.73 $\times ~ 10^{-5}$ & 2.97 $\times ~ 10^{-4}$  \\
$^{174}$Hg &  7.233 & (2.0 $\pm ~ $0.4) $\times ~ 10^{-3}$ & $\minus$0.093\phantom{$\minus$} & 1348.463 & 1348.616  &  0.247  &  0.242 & 2.55 $\times ~ 10^{-3}$ & 5.68 $\times ~ 10^{-4}$ & 2.12 $\times ~ 10^{-3}$  \\
$^{178}$Pb &  7.789 & (2.5 $\pm ~ $0.8) $\times ~ 10^{-4}$ & \phantom{$\minus$}0.000\phantom{$\minus$} & 1368.969 & 1369.559  & ---  &  0.246  & 3.16 $\times ~ 10^{-4}$ & 5.73 $\times ~ 10^{-5}$ & 2.25 $\times ~ 10^{-4}$  \\
$^{180}$Pb &  7.419 & (4.1 $\pm ~ $0.3) $\times ~ 10^{-3}$ & \phantom{$\minus$}0.000\phantom{$\minus$} & 1390.626 & 1390.864  &  0.248  &  0.250 & 3.98 $\times ~ 10^{-3}$ & 8.46 $\times ~ 10^{-4}$ & 2.69 $\times ~ 10^{-3}$  \\
$^{182}$Pb &  7.066 & (5.5 $\pm ~ $0.5) $\times ~ 10^{-2}$ & \phantom{$\minus$}0.000\phantom{$\minus$} & 1411.652 & 1412.208  &  0.234  &  0.289 & 4.68 $\times ~ 10^{-2}$ & 1.34 $\times ~ 10^{-2}$ & 3.48 $\times ~ 10^{-2}$  \\
$^{186}$Po &  8.501 & (3.4 $\pm ~ $1.2) $\times ~ 10^{-5}$ & \phantom{$\minus$}0.327\phantom{$\minus$} & 1431.446 & 1431.001  & ---  &  0.311  & 1.20 $\times ~ 10^{-5}$ & 2.21 $\times ~ 10^{-6}$ & 9.89 $\times ~ 10^{-6}$  \\
$^{188}$Po &  8.082 & (2.7 $\pm ~ $0.3) $\times ~ 10^{-4}$ & \phantom{$\minus$}0.313\phantom{$\minus$} & 1452.235 & 1452.182  &  0.275  &  0.306 & 1.65 $\times ~ 10^{-4}$ & 3.43 $\times ~ 10^{-5}$ & 1.21 $\times ~ 10^{-4}$  \\
$^{190}$Po &  7.693 & (2.45 $\pm ~ $0.05) $\times ~ 10^{-3}$ & $\minus$0.212\phantom{$\minus$} & 1472.396 & 1472.482  &  0.262  &  0.251 & 1.70 $\times ~ 10^{-3}$ & 5.34 $\times ~ 10^{-4}$ & 1.51 $\times ~ 10^{-3}$  \\
$^{192}$Po &  7.320 & (3.22 $\pm ~ $0.03) $\times ~ 10^{-2}$ & $\minus$0.204\phantom{$\minus$} & 1492.042 & 1491.907  &  0.257  &  0.244 & 2.68 $\times ~ 10^{-2}$ & 9.20 $\times ~ 10^{-3}$ & 2.10 $\times ~ 10^{-2}$  \\
$^{194}$Po &  6.987 & (3.92 $\pm ~ $0.04) $\times ~ 10^{-1}$ & $\minus$0.192\phantom{$\minus$} & 1511.123 & 1510.906  &  0.235  &  0.232 & 3.93 $\times ~ 10^{-1}$ & 1.40 $\times ~ 10^{-1}$ & 2.62 $\times ~ 10^{-1}$  \\
$^{208}$Po &  5.216 & (9.145 $\pm ~ $0.006) $\times ~ 10^{7}$ & \phantom{$\minus$}0.000\phantom{$\minus$} & 1630.586 & 1631.017  &  0.135  &  0.152 & 7.16 $\times ~ 10^{7}$ & 1.76 $\times ~ 10^{7}$ & 1.15 $\times ~ 10^{7}$  \\
$^{210}$Po &  5.408 & (1.19557 $\pm ~ $0.00002) $\times ~ 10^{7}$ & \phantom{$\minus$}0.000\phantom{$\minus$} & 1645.213 & 1644.833  &  0.105  &  0.087 & 9.59 $\times ~ 10^{6}$ & 1.30 $\times ~ 10^{6}$ & 9.37 $\times ~ 10^{5}$  \\
$^{212}$Po &  8.954 & (2.944 $\pm ~ $0.008) $\times ~ 10^{-7}$ & \phantom{$\minus$}0.000\phantom{$\minus$} & 1655.772 & 1656.192  &  0.221  &  0.167 & 2.97 $\times ~ 10^{-7}$ & 4.44 $\times ~ 10^{-8}$ & 2.52 $\times ~ 10^{-7}$  \\
$^{214}$Po &  7.834 & (1.6347 $\pm ~ $0.0003) $\times ~ 10^{-4}$ & \phantom{$\minus$}0.000\phantom{$\minus$} & 1666.015 & 1665.734  &  0.213  &  0.142 & 3.59 $\times ~ 10^{-4}$ & 7.11 $\times ~ 10^{-5}$ & 2.06 $\times ~ 10^{-4}$  \\
$^{216}$Po &  6.906 & (1.440 $\pm ~ $0.006) $\times ~ 10^{-1}$ & \phantom{$\minus$}0.089\phantom{$\minus$} & 1675.905 & 1675.435  &  0.205  &  0.122 & 4.89 $\times ~ 10^{-1}$ & 1.19 $\times ~ 10^{-1}$ & 1.99 $\times ~ 10^{-1}$  \\
$^{218}$Po &  6.115 & (1.858 $\pm ~ $0.007) $\times ~ 10^{2}$ & \phantom{$\minus$}0.107\phantom{$\minus$} & 1685.474 & 1685.046  &  0.196  &  0.111 & 8.26 $\times ~ 10^{2}$ & 2.45 $\times ~ 10^{2}$ & 2.57 $\times ~ 10^{2}$  \\
$^{194}$Rn &  7.862 & (7.8 $\pm ~ $1.6) $\times ~ 10^{-4}$ & \phantom{$\minus$}0.166\phantom{$\minus$} & 1492.829 & 1492.661  &  0.261  &  0.249 & 2.57 $\times ~ 10^{-3}$ & 9.70 $\times ~ 10^{-4}$ & 2.34 $\times ~ 10^{-3}$  \\
$^{196}$Rn &  7.617 & (4.7 $\pm ~ $1.1) $\times ~ 10^{-3}$ & \phantom{$\minus$}0.155\phantom{$\minus$} & 1512.721 & 1512.443  &  0.256  &  0.236 & 1.58 $\times ~ 10^{-2}$ & 5.88 $\times ~ 10^{-3}$ & 1.23 $\times ~ 10^{-2}$  \\
$^{212}$Rn &  6.385 & (1.43 $\pm ~ $0.07) $\times ~ 10^{3}$ & \phantom{$\minus$}0.000\phantom{$\minus$} & 1652.497 & 1652.796  &  0.121  &  0.227 & 6.13 $\times ~ 10^{2}$ & 1.92 $\times ~ 10^{2}$ & 1.90 $\times ~ 10^{2}$  \\
$^{214}$Rn &  9.208 & (2.59 $\pm ~ $0.03) $\times ~ 10^{-7}$ & \phantom{$\minus$}0.000\phantom{$\minus$} & 1664.300 & 1665.136  &  0.228  &  0.224 & 2.63 $\times ~ 10^{-7}$ & 5.54 $\times ~ 10^{-8}$ & 2.82 $\times ~ 10^{-7}$  \\
$^{216}$Rn &  8.198 & (2.9 $\pm ~ $0.4) $\times ~ 10^{-5}$ & \phantom{$\minus$}0.060\phantom{$\minus$} & 1675.870 & 1675.719  &  0.237  &  0.197 & 1.25 $\times ~ 10^{-4}$ & 3.48 $\times ~ 10^{-5}$ & 9.71 $\times ~ 10^{-5}$  \\
$^{218}$Rn &  7.262 & (3.375 $\pm ~ $0.015) $\times ~ 10^{-2}$ & \phantom{$\minus$}0.106\phantom{$\minus$} & 1687.048 & 1687.230  &  0.234  &  0.248 & 8.65 $\times ~ 10^{-2}$ & 4.36 $\times ~ 10^{-2}$ & 7.01 $\times ~ 10^{-2}$  \\
$^{220}$Rn &  6.405 & (5.56 $\pm ~ $0.01) $\times ~ 10^{1}$ & \phantom{$\minus$}0.126\phantom{$\minus$} & 1697.796 & 1697.858  &  0.221  &  0.196 & 1.50 $\times ~ 10^{2}$ & 1.16 $\times ~ 10^{2}$ & 1.12 $\times ~ 10^{2}$  \\
$^{222}$Rn &  5.590 & (3.3018 $\pm ~ $0.0002) $\times ~ 10^{5}$ & \phantom{$\minus$}0.142\phantom{$\minus$} & 1708.179 & 1708.144  &  0.222  &  0.185 & 9.37 $\times ~ 10^{5}$ & 1.06 $\times ~ 10^{6}$ & 6.35 $\times ~ 10^{5}$  \\
$^{202}$Ra &  7.880 & (4.1 $\pm ~ $1.1) $\times ~ 10^{-3}$ & \phantom{$\minus$}0.144\phantom{$\minus$} & 1552.485 & 1551.693  &  0.248  &  0.181 & 1.89 $\times ~ 10^{-2}$ & 4.61 $\times ~ 10^{-3}$ & 8.64 $\times ~ 10^{-3}$  \\
$^{204}$Ra &  7.637 & (6 $\pm ~ $0.9) $\times ~ 10^{-2}$ & \phantom{$\minus$}0.131\phantom{$\minus$} & 1571.641 & 1571.806  &  0.231  &  0.256 & 8.14 $\times ~ 10^{-2}$ & 2.86 $\times ~ 10^{-2}$ & 4.65 $\times ~ 10^{-2}$  \\
$^{206}$Ra &  7.415 & (2.4 $\pm ~ $0.2) $\times ~ 10^{-1}$ & \phantom{$\minus$}0.115\phantom{$\minus$} & 1590.279 & 1590.252  &  0.219  &  0.207 & 5.59 $\times ~ 10^{-1}$ & 1.62 $\times ~ 10^{-1}$ & 2.31 $\times ~ 10^{-1}$  \\
$^{210}$Ra &  7.151 & (4.0 $\pm ~ $0.1) $\times ~ 10^{0}$ & \phantom{$\minus$}0.049\phantom{$\minus$} & 1625.687 & 1626.120  &  0.187  &  0.255 & 3.90 $\times ~ 10^{0}$ & 1.34 $\times ~ 10^{0}$ & 1.63 $\times ~ 10^{0}$  \\
$^{214}$Ra &  7.273 & (2.437 $\pm ~ $0.016) $\times ~ 10^{0}$ & \phantom{$\minus$}0.000\phantom{$\minus$} & 1658.323 & 1658.615  &  0.138  &  0.204 & 1.58 $\times ~ 10^{0}$ & 3.95 $\times ~ 10^{-1}$ & 5.08 $\times ~ 10^{-1}$  \\
$^{216}$Ra &  9.526 & (1.72 $\pm ~ $0.07) $\times ~ 10^{-7}$ & \phantom{$\minus$}0.000\phantom{$\minus$} & 1671.267 & 1671.962  &  0.239  &  0.228 & 2.19 $\times ~ 10^{-7}$ & 4.79 $\times ~ 10^{-8}$ & 2.27 $\times ~ 10^{-7}$  \\
$^{218}$Ra &  8.540 & (2.591 $\pm ~ $0.014) $\times ~ 10^{-5}$ & \phantom{$\minus$}0.059\phantom{$\minus$} & 1684.055 & 1684.379  &  0.243  &  0.264 & 5.53 $\times ~ 10^{-5}$ & 2.09 $\times ~ 10^{-5}$ & 5.53 $\times ~ 10^{-5}$  \\
$^{220}$Ra &  7.594 & (1.81 $\pm ~ $0.12) $\times ~ 10^{-2}$ & \phantom{$\minus$}0.093\phantom{$\minus$} & 1696.571 & 1696.703  &  0.239  &  0.210 & 4.61 $\times ~ 10^{-2}$ & 2.13 $\times ~ 10^{-2}$ & 3.25 $\times ~ 10^{-2}$  \\
$^{222}$Ra &  6.678 & (3.36 $\pm ~ $0.04) $\times ~ 10^{1}$ & \phantom{$\minus$}0.122\phantom{$\minus$} & 1708.666 & 1709.011  &  0.199  &  0.209 & 7.94 $\times ~ 10^{1}$ & 6.89 $\times ~ 10^{1}$ & 6.14 $\times ~ 10^{1}$  \\
$^{224}$Ra &  5.789 & (3.1377 $\pm ~ $0.0012) $\times ~ 10^{5}$ & \phantom{$\minus$}0.151\phantom{$\minus$} & 1720.303 & 1720.964  &  0.184  &  0.187 & 8.38 $\times ~ 10^{5}$ & 1.04 $\times ~ 10^{6}$ & 5.54 $\times ~ 10^{5}$  \\
$^{226}$Ra &  4.871 & (5.049 $\pm ~ $0.02) $\times ~ 10^{10}$ & \phantom{$\minus$}0.187\phantom{$\minus$} & 1731.604 & 1731.951  &  0.182  &  0.158 & 1.98 $\times ~ 10^{11}$ & 3.16 $\times ~ 10^{11}$ & 9.84 $\times ~ 10^{10}$  \\
$^{208}$Th &  8.202 & (2.4 $\pm ~ $1.2) $\times ~ 10^{-3}$ & \phantom{$\minus$}0.118\phantom{$\minus$} & 1591.735 & 1591.922  & ---  &  0.251  & 7.38 $\times ~ 10^{-3}$ & 2.39 $\times ~ 10^{-3}$ & 4.16 $\times ~ 10^{-3}$  \\
$^{210}$Th &  8.069 & (1.60 $\pm ~ $0.36) $\times ~ 10^{-2}$ & \phantom{$\minus$}0.085\phantom{$\minus$} & 1610.506 & 1610.464  &  0.234  &  0.227 & 2.04 $\times ~ 10^{-2}$ & 5.94 $\times ~ 10^{-3}$ & 9.55 $\times ~ 10^{-3}$  \\
$^{212}$Th &  7.958 & (3.17 $\pm ~ $0.13) $\times ~ 10^{-2}$ & \phantom{$\minus$}0.040\phantom{$\minus$} & 1628.597 & 1628.480  &  0.196  &  0.227 & 4.61 $\times ~ 10^{-2}$ & 1.28 $\times ~ 10^{-2}$ & 1.92 $\times ~ 10^{-2}$  \\
$^{214}$Th &  7.827 & (8.7 $\pm ~ $1) $\times ~ 10^{-2}$ & \phantom{$\minus$}0.000\phantom{$\minus$} & 1646.156 & 1646.998  &  0.195  &  0.276 & 9.85 $\times ~ 10^{-2}$ & 3.28 $\times ~ 10^{-2}$ & 4.55 $\times ~ 10^{-2}$  \\
$^{216}$Th &  8.072 & (2.628 $\pm ~ $0.016) $\times ~ 10^{-2}$ & \phantom{$\minus$}0.000\phantom{$\minus$} & 1662.695 & 1662.952  &  0.159  &  0.209 & 2.00 $\times ~ 10^{-2}$ & 4.55 $\times ~ 10^{-3}$ & 7.22 $\times ~ 10^{-3}$  \\
$^{218}$Th &  9.849 & (1.22 $\pm ~ $0.05) $\times ~ 10^{-7}$ & \phantom{$\minus$}0.000\phantom{$\minus$} & 1676.769 & 1677.516  &  0.251  &  0.256 & 1.63 $\times ~ 10^{-7}$ & 4.07 $\times ~ 10^{-8}$ & 1.80 $\times ~ 10^{-7}$  \\
$^{220}$Th &  8.973 & (1.02 $\pm ~ $0.03) $\times ~ 10^{-5}$ & \phantom{$\minus$}0.018\phantom{$\minus$} & 1690.589 & 1690.809  &  0.243  &  0.229 & 2.28 $\times ~ 10^{-5}$ & 7.25 $\times ~ 10^{-6}$ & 1.92 $\times ~ 10^{-5}$  \\
$^{222}$Th &  8.133 & (2.24 $\pm ~ $0.03) $\times ~ 10^{-3}$ & \phantom{$\minus$}0.067\phantom{$\minus$} & 1704.218 & 1704.640  &  0.232  &  0.249 & 4.50 $\times ~ 10^{-3}$ & 2.28 $\times ~ 10^{-3}$ & 3.71 $\times ~ 10^{-3}$  \\
%\hline
 \hline
  \end{tabular}\\
     %\end{center}
     \end{adjustwidth}
\end{table*}
%%%%%%%%%%%%%%%%

%%%%%%%%%%%%%%%
\begin{table*}[b]
%\begin{center}
 \caption{$(Continued.)$
\label{hl_tbl2} }
 \begin{adjustwidth}{-0.5cm}{}
   \begin{tabular}{c|c|l|c|c|c|c|c|l|l|l}
    \hline
 \multirow{2}{*}{\parbox{1cm}{Parent \\nucleus}} & \multicolumn{2}{c|}{Experimental data} & \multirow{2}{*}{\parbox{1.2cm}{$\beta_{2,{\rm p}}$ \\ (DRHBc)}} & \multicolumn{2}{c|}{ $E_{\rm b}$ } & \multicolumn{2}{c|}{Preformation factor} & \multicolumn{3}{c}{ $T^{\rm cal}_{1/2}$ }  \\  
 \cline{2-3}
 \cline{5-11}
  & \phantom{a } $\Qae$ \phantom{a } & \multicolumn{1}{c|}{$T^{\rm exp}_{1/2}$} & & AME2020 & DNN study & AME2020 & DNN study & \multicolumn{1}{c|}{Our result} & \multicolumn{1}{c|}{ZZCW} & \multicolumn{1}{c}{UNIV} \\
    \hline
$^{224}$Th &  7.299 & (1.04 $\pm ~ $0.02) $\times ~ 10^{0}$ & \phantom{$\minus$}0.081\phantom{$\minus$} & 1717.568 & 1717.850  &  0.198  &  0.204 & 2.83 $\times ~ 10^{0}$ & 1.81 $\times ~ 10^{0}$ & 1.81 $\times ~ 10^{0}$  \\
$^{226}$Th &  6.453 & (1.842 $\pm ~ $0.002) $\times ~ 10^{3}$ & \phantom{$\minus$}0.184\phantom{$\minus$} & 1730.509 & 1730.909  &  0.182  &  0.188 & 5.84 $\times ~ 10^{3}$ & 5.81 $\times ~ 10^{3}$ & 3.56 $\times ~ 10^{3}$  \\
$^{228}$Th &  5.520 & (3.1951 $\pm ~ $0.002) $\times ~ 10^{7}$ & \phantom{$\minus$}0.227\phantom{$\minus$} & 1743.078 & 1743.705  &  0.183  &  0.213 & 1.70 $\times ~ 10^{8}$ & 3.47 $\times ~ 10^{8}$ & 1.24 $\times ~ 10^{8}$  \\
$^{230}$Th &  4.770 & (2.38 $\pm ~ $0.01) $\times ~ 10^{12}$ & \phantom{$\minus$}0.243\phantom{$\minus$} & 1755.129 & 1755.316  &  0.183  &  0.176 & 6.62 $\times ~ 10^{12}$ & 2.30 $\times ~ 10^{13}$ & 5.31 $\times ~ 10^{12}$  \\
$^{232}$Th &  4.082 & (4.40 $\pm ~ $0.3) $\times ~ 10^{16}$ & \phantom{$\minus$}0.263\phantom{$\minus$} & 1766.688 & 1766.413  &  0.160  &  0.140 & 1.56 $\times ~ 10^{18}$ & 8.25 $\times ~ 10^{18}$ & 1.27 $\times ~ 10^{18}$  \\
$^{216}$U &  8.531 & (6.9 $\pm ~ $2.9) $\times ~ 10^{-3}$ & \phantom{$\minus$}0.000\phantom{$\minus$} & 1648.362 & 1649.859  &  0.206  &  0.225 & 4.73 $\times ~ 10^{-3}$ & 1.13 $\times ~ 10^{-3}$ & 1.83 $\times ~ 10^{-3}$  \\
$^{218}$U &  8.775 & (3.54 $\pm ~ $0.91) $\times ~ 10^{-4}$ & \phantom{$\minus$}0.000\phantom{$\minus$} & 1665.677 & 1666.961  &  0.183  &  0.234 & 8.62 $\times ~ 10^{-4}$ & 1.90 $\times ~ 10^{-4}$ & 3.53 $\times ~ 10^{-4}$  \\
$^{222}$U &  9.481 & (4.7 $\pm ~ $0.7) $\times ~ 10^{-6}$ & \phantom{$\minus$}0.000\phantom{$\minus$} & 1695.584 & 1696.134  &  0.245  &  0.238 & 5.57 $\times ~ 10^{-6}$ & 1.73 $\times ~ 10^{-6}$ & 4.79 $\times ~ 10^{-6}$  \\
$^{224}$U &  8.628 & (3.96 $\pm ~ $0.17) $\times ~ 10^{-4}$ & \phantom{$\minus$}0.000\phantom{$\minus$} & 1710.257 & 1710.541  &  0.247  &  0.210 & 1.08 $\times ~ 10^{-3}$ & 4.13 $\times ~ 10^{-4}$ & 6.97 $\times ~ 10^{-4}$  \\
$^{226}$U &  7.701 & (2.69 $\pm ~ $0.06) $\times ~ 10^{-1}$ & \phantom{$\minus$}0.000\phantom{$\minus$} & 1724.813 & 1724.947  &  0.207  &  0.221 & 7.27 $\times ~ 10^{-1}$ & 4.44 $\times ~ 10^{-1}$ & 4.38 $\times ~ 10^{-1}$  \\
$^{230}$U &  5.992 & (1.748 $\pm ~ $0.002) $\times ~ 10^{6}$ & \phantom{$\minus$}0.246\phantom{$\minus$} & 1752.812 & 1752.545  &  0.187  &  0.184 & 3.61 $\times ~ 10^{6}$ & 9.83 $\times ~ 10^{6}$ & 3.63 $\times ~ 10^{6}$  \\
$^{232}$U &  5.414 & (2.17 $\pm ~ $0.01) $\times ~ 10^{9}$ & \phantom{$\minus$}0.258\phantom{$\minus$} & 1765.960 & 1765.870  &  0.169  &  0.200 & 2.37 $\times ~ 10^{9}$ & 1.73 $\times ~ 10^{10}$ & 4.55 $\times ~ 10^{9}$  \\
$^{234}$U &  4.858 & (7.747 $\pm ~ $0.02) $\times ~ 10^{12}$ & \phantom{$\minus$}0.276\phantom{$\minus$} & 1778.568 & 1778.714  &  0.150  &  0.183 & 9.06 $\times ~ 10^{12}$ & 7.76 $\times ~ 10^{13}$ & 1.48 $\times ~ 10^{13}$  \\
$^{236}$U &  4.573 & (7.391 $\pm ~ $0.01) $\times ~ 10^{14}$ & \phantom{$\minus$}0.290\phantom{$\minus$} & 1790.411 & 1790.047  &  0.153  &  0.131 & 1.34 $\times ~ 10^{15}$ & 9.99 $\times ~ 10^{15}$ & 1.60 $\times ~ 10^{15}$  \\
$^{238}$U &  4.270 & (1.408 $\pm ~ $0.001) $\times ~ 10^{17}$ & \phantom{$\minus$}0.288\phantom{$\minus$} & 1801.690 & 1800.749  &  0.137  &  0.104 & 4.26 $\times ~ 10^{17}$ & 3.02 $\times ~ 10^{18}$ & 4.04 $\times ~ 10^{17}$  \\
   \hline
    \hline
  \end{tabular}\\
    % \end{center}
     \end{adjustwidth}
\end{table*}
%%%%%%%%%%%%%%%%

%%%%%%%%%%%%%%%
\begin{table*}[b]
\begin{center}
 \caption{The predicted half-lives with our method, ZZCW, and UNIV. The experimental data $\Qae$ for the DRHBc, ZZCW, and UNIV is taken from AME2020~\cite{Wang:2021xhn}. The units of $\Qae$ and $T^{\rm cal}_{1/2}$ are MeV and s, respectively. 
\label{hl_tbl_pd} }
{\small
  \begin{tabular}{c|c|c|l|l|l}
   \hline
    \hline
 \multirow{2}{*}{\parbox{1cm}{Parent \\nucleus}} & \multirow{2}{*}{ \phantom{a } $\Qae$ \phantom{a } } & \multicolumn{2}{c|}{ Our result } & \multicolumn{1}{c|}{ZZCW} & \multicolumn{1}{c}{UNIV} \\
 \cline{3-6}
 & & \phantom{a } $\Pa$ \phantom{a } & \multicolumn{1}{c|}{$T^{\rm cal}_{1/2}$} & \multicolumn{1}{c|}{$T^{\rm cal}_{1/2}$} & \multicolumn{1}{c}{$T^{\rm cal}_{1/2}$} \\
   \hline
$^{160}$W &  6.066 &  0.273 & 1.39 $\times ~ 10^{-1}$ & 1.02 $\times ~ 10^{-1}$ & 1.27 $\times ~ 10^{-1}$ \\
$^{162}$W &  5.678 &  0.259 & 4.33 $\times ~ 10^{0}$ & 3.19 $\times ~ 10^{0}$ & 4.36 $\times ~ 10^{0}$ \\
$^{164}$W &  5.278 &  0.244 & 2.59 $\times ~ 10^{2}$ & 1.77 $\times ~ 10^{2}$ & 2.62 $\times ~ 10^{2}$ \\
$^{166}$W &  4.856 &  0.255 & 3.51 $\times ~ 10^{4}$ & 2.27 $\times ~ 10^{4}$ & 3.54 $\times ~ 10^{4}$ \\
$^{168}$W &  4.501 &  0.202 & 4.90 $\times ~ 10^{6}$ & 2.37 $\times ~ 10^{6}$ & 3.81 $\times ~ 10^{6}$ \\
$^{170}$W &  4.143 &  0.184 & 1.09 $\times ~ 10^{9}$ & 4.86 $\times ~ 10^{8}$ & 7.89 $\times ~ 10^{8}$ \\
$^{172}$W &  3.838 &  0.186 & 1.58 $\times ~ 10^{11}$ & 8.48 $\times ~ 10^{10}$ & 1.36 $\times ~ 10^{11}$ \\
$^{174}$W &  3.602 &  0.219 & 8.00 $\times ~ 10^{12}$ & 7.49 $\times ~ 10^{12}$ & 1.15 $\times ~ 10^{13}$ \\
$^{176}$W &  3.336 &  0.171 & 2.94 $\times ~ 10^{15}$ & 2.13 $\times ~ 10^{15}$ & 3.07 $\times ~ 10^{15}$ \\
$^{178}$W &  3.013 &  0.143 & 1.13 $\times ~ 10^{19}$ & 5.71 $\times ~ 10^{18}$ & 7.65 $\times ~ 10^{18}$ \\
$^{182}$W &  1.764 &  0.171 & 6.30 $\times ~ 10^{40}$ & 1.37 $\times ~ 10^{40}$ & 1.62 $\times ~ 10^{40}$ \\
$^{184}$W &  1.649 &  0.157 & 1.04 $\times ~ 10^{44}$ & 2.04 $\times ~ 10^{43}$ & 2.09 $\times ~ 10^{43}$ \\
$^{186}$W &  1.116 &  0.158 & 5.96 $\times ~ 10^{64}$ & 6.08 $\times ~ 10^{63}$ & 5.93 $\times ~ 10^{63}$ \\
$^{164}$Os &  6.479 &  0.266 & 2.39 $\times ~ 10^{-2}$ & 2.08 $\times ~ 10^{-2}$ & 2.40 $\times ~ 10^{-2}$ \\
$^{166}$Os &  6.143 &  0.263 & 3.79 $\times ~ 10^{-1}$ & 3.18 $\times ~ 10^{-1}$ & 3.96 $\times ~ 10^{-1}$ \\
$^{168}$Os &  5.816 &  0.240 & 8.18 $\times ~ 10^{0}$ & 5.93 $\times ~ 10^{0}$ & 7.76 $\times ~ 10^{0}$ \\
$^{170}$Os &  5.537 &  0.238 & 1.29 $\times ~ 10^{2}$ & 9.02 $\times ~ 10^{1}$ & 1.21 $\times ~ 10^{2}$ \\
$^{172}$Os &  5.224 &  0.241 & 3.66 $\times ~ 10^{3}$ & 2.61 $\times ~ 10^{3}$ & 3.53 $\times ~ 10^{3}$ \\
$^{174}$Os &  4.871 &  0.210 & 1.45 $\times ~ 10^{5}$ & 1.85 $\times ~ 10^{5}$ & 2.47 $\times ~ 10^{5}$ \\
$^{176}$Os &  4.541 &  0.197 & 1.37 $\times ~ 10^{7}$ & 1.59 $\times ~ 10^{7}$ & 2.04 $\times ~ 10^{7}$ \\
$^{178}$Os &  4.258 &  0.209 & 9.37 $\times ~ 10^{8}$ & 1.14 $\times ~ 10^{9}$ & 1.38 $\times ~ 10^{9}$ \\
$^{180}$Os &  3.860 &  0.194 & 1.05 $\times ~ 10^{12}$ & 1.07 $\times ~ 10^{12}$ & 1.19 $\times ~ 10^{12}$ \\
$^{182}$Os &  3.373 &  0.185 & 3.38 $\times ~ 10^{16}$ & 2.49 $\times ~ 10^{16}$ & 2.54 $\times ~ 10^{16}$ \\
$^{184}$Os &  2.959 &  0.165 & 1.80 $\times ~ 10^{21}$ & 9.15 $\times ~ 10^{20}$ & 8.35 $\times ~ 10^{20}$ \\
$^{188}$Os &  2.143 &  0.176 & 1.25 $\times ~ 10^{34}$ & 4.45 $\times ~ 10^{33}$ & 3.10 $\times ~ 10^{33}$ \\
$^{190}$Os &  1.376 &  0.190 & 3.74 $\times ~ 10^{55}$ & 7.22 $\times ~ 10^{54}$ & 4.69 $\times ~ 10^{54}$ \\
$^{172}$Pt &  6.463 &  0.235 & 1.62 $\times ~ 10^{-1}$ & 1.31 $\times ~ 10^{-1}$ & 1.44 $\times ~ 10^{-1}$ \\
$^{174}$Pt &  6.183 &  0.260 & 1.63 $\times ~ 10^{0}$ & 1.41 $\times ~ 10^{0}$ & 1.55 $\times ~ 10^{0}$ \\
$^{176}$Pt &  5.885 &  0.245 & 1.16 $\times ~ 10^{1}$ & 2.24 $\times ~ 10^{1}$ & 2.42 $\times ~ 10^{1}$ \\
$^{178}$Pt &  5.573 &  0.206 & 3.18 $\times ~ 10^{2}$ & 5.32 $\times ~ 10^{2}$ & 5.54 $\times ~ 10^{2}$ \\
$^{180}$Pt &  5.276 &  0.222 & 8.09 $\times ~ 10^{3}$ & 1.46 $\times ~ 10^{4}$ & 1.43 $\times ~ 10^{4}$ \\
$^{182}$Pt &  4.951 &  0.201 & 5.15 $\times ~ 10^{5}$ & 8.07 $\times ~ 10^{5}$ & 7.28 $\times ~ 10^{5}$ \\
$^{184}$Pt &  4.599 &  0.193 & 7.96 $\times ~ 10^{7}$ & 1.04 $\times ~ 10^{8}$ & 8.46 $\times ~ 10^{7}$ \\
$^{186}$Pt &  4.320 &  0.212 & 5.97 $\times ~ 10^{9}$ & 7.73 $\times ~ 10^{9}$ & 5.54 $\times ~ 10^{9}$ \\
$^{188}$Pt &  4.007 &  0.195 & 1.56 $\times ~ 10^{12}$ & 1.69 $\times ~ 10^{12}$ & 1.05 $\times ~ 10^{12}$ \\
$^{192}$Pt &  2.424 &  0.178 & 4.96 $\times ~ 10^{30}$ & 2.54 $\times ~ 10^{30}$ & 1.19 $\times ~ 10^{30}$ \\
$^{194}$Pt &  1.523 &  0.160 & 5.56 $\times ~ 10^{52}$ & 9.82 $\times ~ 10^{51}$ & 4.21 $\times ~ 10^{51}$ \\
$^{196}$Pt &  0.813 &  0.168 & 7.85 $\times ~ 10^{91}$ & 3.52 $\times ~ 10^{90}$ & 1.53 $\times ~ 10^{90}$ \\
$^{206}$Pt &  0.865 &  0.127 & 2.07 $\times ~ 10^{87}$ & 3.35 $\times ~ 10^{86}$ & 4.22 $\times ~ 10^{85}$ \\
$^{176}$Hg &  6.897 &  0.234 & 3.25 $\times ~ 10^{-2}$ & 2.81 $\times ~ 10^{-2}$ & 2.49 $\times ~ 10^{-2}$ \\
$^{178}$Hg &  6.577 &  0.286 & 1.35 $\times ~ 10^{-1}$ & 3.59 $\times ~ 10^{-1}$ & 3.13 $\times ~ 10^{-1}$ \\
$^{180}$Hg &  6.258 &  0.262 & 2.19 $\times ~ 10^{0}$ & 5.76 $\times ~ 10^{0}$ & 4.84 $\times ~ 10^{0}$ \\
$^{182}$Hg &  5.996 &  0.236 & 2.86 $\times ~ 10^{1}$ & 6.88 $\times ~ 10^{1}$ & 5.43 $\times ~ 10^{1}$ \\
$^{184}$Hg &  5.660 &  0.234 & 9.62 $\times ~ 10^{2}$ & 2.19 $\times ~ 10^{3}$ & 1.59 $\times ~ 10^{3}$ \\
$^{186}$Hg &  5.204 &  0.234 & 2.12 $\times ~ 10^{5}$ & 4.28 $\times ~ 10^{5}$ & 2.82 $\times ~ 10^{5}$ \\
$^{188}$Hg &  4.709 &  0.222 & 1.98 $\times ~ 10^{8}$ & 3.28 $\times ~ 10^{8}$ & 1.92 $\times ~ 10^{8}$ \\
$^{190}$Hg &  4.069 &  0.176 & 9.48 $\times ~ 10^{12}$ & 1.07 $\times ~ 10^{13}$ & 5.51 $\times ~ 10^{12}$ \\
$^{192}$Hg &  3.385 &  0.179 & 1.98 $\times ~ 10^{19}$ & 1.80 $\times ~ 10^{19}$ & 8.03 $\times ~ 10^{18}$ \\
$^{194}$Hg &  2.698 &  0.178 & 1.78 $\times ~ 10^{28}$ & 6.50 $\times ~ 10^{27}$ & 2.49 $\times ~ 10^{27}$ \\
$^{196}$Hg &  2.038 &  0.175 & 3.05 $\times ~ 10^{40}$ & 7.84 $\times ~ 10^{39}$ & 2.57 $\times ~ 10^{39}$ \\
$^{198}$Hg &  1.381 &  0.175 & 4.38 $\times ~ 10^{60}$ & 6.15 $\times ~ 10^{59}$ & 1.78 $\times ~ 10^{59}$ \\
$^{208}$Hg &  1.930 &  0.141 & 6.74 $\times ~ 10^{42}$ & 6.90 $\times ~ 10^{42}$ & 5.35 $\times ~ 10^{41}$ \\
$^{210}$Hg &  1.515 &  0.132 & 1.67 $\times ~ 10^{55}$ & 1.02 $\times ~ 10^{55}$ & 8.59 $\times ~ 10^{53}$ \\
$^{212}$Hg &  1.095 &  0.114 & 3.04 $\times ~ 10^{74}$ & 7.97 $\times ~ 10^{73}$ & 7.55 $\times ~ 10^{72}$ \\
    \hline
%     \hline
  \end{tabular}\\
}
     \end{center}
\end{table*}
%%%%%%%%%%%%%%%%

\begin{table*}[b]
\begin{center}
 \caption{$(Continued.)$ 
\label{hl_tbl_pd2}  }
{\small
  \begin{tabular}{c|c|c|l|l|l}
    \hline
 \multirow{2}{*}{\parbox{1cm}{Parent \\nucleus}} & \multirow{2}{*}{ \phantom{a } $\Qae$ \phantom{a } } & \multicolumn{2}{c|}{ Our result } & \multicolumn{1}{c|}{ZZCW} & \multicolumn{1}{c}{UNIV} \\
 \cline{3-6}
 & & \phantom{a } $\Pa$ \phantom{a } & \multicolumn{1}{c|}{$T^{\rm cal}_{1/2}$} & \multicolumn{1}{c|}{$T^{\rm cal}_{1/2}$} & \multicolumn{1}{c}{$T^{\rm cal}_{1/2}$} \\
   \hline
$^{184}$Pb &  6.774 &  0.265 & 1.56 $\times ~ 10^{-1}$ & 7.92 $\times ~ 10^{-1}$ & 3.35 $\times ~ 10^{-1}$ \\
$^{186}$Pb &  6.471 &  0.238 & 2.53 $\times ~ 10^{0}$ & 9.16 $\times ~ 10^{0}$ & 4.23 $\times ~ 10^{0}$ \\
$^{188}$Pb &  6.109 &  0.253 & 7.73 $\times ~ 10^{1}$ & 2.43 $\times ~ 10^{2}$ & 1.17 $\times ~ 10^{2}$ \\
$^{190}$Pb &  5.698 &  0.236 & 6.79 $\times ~ 10^{3}$ & 1.61 $\times ~ 10^{4}$ & 7.75 $\times ~ 10^{3}$ \\
$^{192}$Pb &  5.222 &  0.194 & 2.50 $\times ~ 10^{6}$ & 4.19 $\times ~ 10^{6}$ & 1.93 $\times ~ 10^{6}$ \\
$^{194}$Pb &  4.738 &  0.201 & 4.41 $\times ~ 10^{9}$ & 3.05 $\times ~ 10^{9}$ & 1.28 $\times ~ 10^{9}$ \\
$^{196}$Pb &  4.238 &  0.207 & 1.41 $\times ~ 10^{13}$ & 9.51 $\times ~ 10^{12}$ & 3.50 $\times ~ 10^{12}$ \\
$^{198}$Pb &  3.692 &  0.156 & 8.85 $\times ~ 10^{17}$ & 4.25 $\times ~ 10^{17}$ & 1.31 $\times ~ 10^{17}$ \\
$^{200}$Pb &  3.150 &  0.146 & 6.27 $\times ~ 10^{23}$ & 2.65 $\times ~ 10^{23}$ & 6.59 $\times ~ 10^{22}$ \\
$^{202}$Pb &  2.589 &  0.143 & 5.40 $\times ~ 10^{31}$ & 1.98 $\times ~ 10^{31}$ & 3.83 $\times ~ 10^{30}$ \\
$^{204}$Pb &  1.969 &  0.126 & 1.97 $\times ~ 10^{44}$ & 4.92 $\times ~ 10^{43}$ & 7.26 $\times ~ 10^{42}$ \\
$^{206}$Pb &  1.135 &  0.118 & 2.85 $\times ~ 10^{75}$ & 2.59 $\times ~ 10^{74}$ & 3.16 $\times ~ 10^{73}$ \\
$^{210}$Pb &  3.792 &  0.111 & 6.14 $\times ~ 10^{16}$ & 1.30 $\times ~ 10^{16}$ & 9.17 $\times ~ 10^{15}$ \\
$^{212}$Pb &  3.292 &  0.098 & 8.14 $\times ~ 10^{21}$ & 8.04 $\times ~ 10^{20}$ & 8.98 $\times ~ 10^{20}$ \\
$^{214}$Pb &  2.692 &  0.115 & 5.61 $\times ~ 10^{29}$ & 3.30 $\times ~ 10^{28}$ & 5.48 $\times ~ 10^{28}$ \\
$^{216}$Pb &  2.065 &  0.095 & 4.24 $\times ~ 10^{41}$ & 9.76 $\times ~ 10^{39}$ & 2.30 $\times ~ 10^{40}$ \\
$^{218}$Pb &  1.434 &  0.079 & 6.26 $\times ~ 10^{60}$ & 4.89 $\times ~ 10^{58}$ & 1.59 $\times ~ 10^{59}$ \\
$^{196}$Po &  6.658 &  0.249 & 6.15 $\times ~ 10^{0}$ & 6.05 $\times ~ 10^{0}$ & 3.91 $\times ~ 10^{0}$ \\
$^{198}$Po &  6.310 &  0.180 & 2.17 $\times ~ 10^{2}$ & 1.58 $\times ~ 10^{2}$ & 8.87 $\times ~ 10^{1}$ \\
$^{200}$Po &  5.982 &  0.200 & 5.40 $\times ~ 10^{3}$ & 4.71 $\times ~ 10^{3}$ & 2.18 $\times ~ 10^{3}$ \\
$^{202}$Po &  5.701 &  0.164 & 1.41 $\times ~ 10^{5}$ & 1.16 $\times ~ 10^{5}$ & 4.21 $\times ~ 10^{4}$ \\
$^{204}$Po &  5.485 &  0.184 & 2.23 $\times ~ 10^{6}$ & 1.75 $\times ~ 10^{6}$ & 4.76 $\times ~ 10^{5}$ \\
$^{206}$Po &  5.327 &  0.164 & 1.68 $\times ~ 10^{7}$ & 1.56 $\times ~ 10^{7}$ & 3.03 $\times ~ 10^{6}$ \\
$^{220}$Po &  5.329 &  0.178 & 3.98 $\times ~ 10^{6}$ & 7.82 $\times ~ 10^{5}$ & 1.63 $\times ~ 10^{6}$ \\
$^{222}$Po &  4.432 &  0.103 & 3.60 $\times ~ 10^{12}$ & 2.65 $\times ~ 10^{11}$ & 6.50 $\times ~ 10^{11}$ \\
$^{224}$Po &  3.355 &  0.088 & 2.83 $\times ~ 10^{22}$ & 1.04 $\times ~ 10^{21}$ & 2.85 $\times ~ 10^{21}$ \\
$^{198}$Rn &  7.349 &  0.211 & 1.37 $\times ~ 10^{-1}$ & 7.93 $\times ~ 10^{-2}$ & 8.37 $\times ~ 10^{-2}$ \\
$^{200}$Rn &  7.043 &  0.242 & 1.43 $\times ~ 10^{0}$ & 9.71 $\times ~ 10^{-1}$ & 8.86 $\times ~ 10^{-1}$ \\
$^{202}$Rn &  6.774 &  0.197 & 1.79 $\times ~ 10^{1}$ & 1.08 $\times ~ 10^{1}$ & 8.11 $\times ~ 10^{0}$ \\
$^{204}$Rn &  6.547 &  0.191 & 1.44 $\times ~ 10^{2}$ & 9.85 $\times ~ 10^{1}$ & 5.80 $\times ~ 10^{1}$ \\
$^{206}$Rn &  6.384 &  0.205 & 8.14 $\times ~ 10^{2}$ & 5.68 $\times ~ 10^{2}$ & 2.50 $\times ~ 10^{2}$ \\
$^{208}$Rn &  6.261 &  0.209 & 2.65 $\times ~ 10^{3}$ & 2.45 $\times ~ 10^{3}$ & 7.66 $\times ~ 10^{2}$ \\
$^{210}$Rn &  6.159 &  0.167 & 9.16 $\times ~ 10^{3}$ & 9.29 $\times ~ 10^{3}$ & 1.97 $\times ~ 10^{3}$ \\
$^{224}$Rn &  4.757 &  0.190 & 7.89 $\times ~ 10^{10}$ & 2.14 $\times ~ 10^{10}$ & 4.65 $\times ~ 10^{10}$ \\
$^{226}$Rn &  3.836 &  0.159 & 1.80 $\times ~ 10^{18}$ & 3.18 $\times ~ 10^{17}$ & 7.61 $\times ~ 10^{17}$ \\
$^{228}$Rn &  2.908 &  0.135 & 7.96 $\times ~ 10^{28}$ & 1.30 $\times ~ 10^{28}$ & 3.23 $\times ~ 10^{28}$ \\
$^{230}$Rn &  2.196 &  0.094 & 3.66 $\times ~ 10^{41}$ & 4.08 $\times ~ 10^{40}$ & 9.91 $\times ~ 10^{40}$ \\
$^{208}$Ra &  7.273 &  0.210 & 1.72 $\times ~ 10^{0}$ & 9.22 $\times ~ 10^{-1}$ & 6.54 $\times ~ 10^{-1}$ \\
$^{212}$Ra &  7.032 &  0.179 & 1.49 $\times ~ 10^{1}$ & 1.19 $\times ~ 10^{1}$ & 4.07 $\times ~ 10^{0}$ \\
$^{228}$Ra &  4.070 &  0.156 & 2.32 $\times ~ 10^{17}$ & 4.90 $\times ~ 10^{16}$ & 1.04 $\times ~ 10^{17}$ \\
$^{230}$Ra &  3.344 &  0.130 & 5.84 $\times ~ 10^{24}$ & 9.80 $\times ~ 10^{23}$ & 2.13 $\times ~ 10^{24}$ \\
$^{232}$Ra &  2.829 &  0.096 & 4.79 $\times ~ 10^{31}$ & 7.56 $\times ~ 10^{30}$ & 1.55 $\times ~ 10^{31}$ \\
$^{234}$Ra &  2.336 &  0.088 & 2.03 $\times ~ 10^{40}$ & 3.59 $\times ~ 10^{39}$ & 6.50 $\times ~ 10^{39}$ \\
$^{234}$Th &  3.672 &  0.135 & 1.26 $\times ~ 10^{22}$ & 6.00 $\times ~ 10^{21}$ & 1.04 $\times ~ 10^{22}$ \\
$^{236}$Th &  3.333 &  0.099 & 9.51 $\times ~ 10^{25}$ & 4.14 $\times ~ 10^{25}$ & 6.15 $\times ~ 10^{25}$ \\
$^{238}$Th &  3.169 &  0.100 & 1.28 $\times ~ 10^{28}$ & 5.70 $\times ~ 10^{27}$ & 6.66 $\times ~ 10^{27}$ \\
$^{220}$U & 10.290 &  0.221 & 8.95 $\times ~ 10^{-8}$ & 2.70 $\times ~ 10^{-7}$ & 8.24 $\times ~ 10^{-8}$ \\
$^{228}$U &  6.800 &  0.219 & 1.15 $\times ~ 10^{3}$ & 6.87 $\times ~ 10^{2}$ & 8.81 $\times ~ 10^{2}$ \\
$^{240}$U &  4.035 &  0.107 & 5.99 $\times ~ 10^{19}$ & 4.46 $\times ~ 10^{19}$ & 4.43 $\times ~ 10^{19}$ \\
$^{242}$U &  3.670 &  0.080 & 4.11 $\times ~ 10^{23}$ & 2.41 $\times ~ 10^{23}$ & 1.76 $\times ~ 10^{23}$ \\
    \hline 
     \hline
   % \hline
  \end{tabular}\\
}
     \end{center}
\end{table*}

\section{Appendix: empirical formulae for $\alpha$-decay half-lives }
\label{sec_appendix}

In the absence of experimental data to compare, our results must be discussed using alternative calculations.
We introduce two recent empirical formulae as tools to simple calculate $\alpha$-decay half-lives and compare our results.

\begin{itemize}
\item {ZZCW}:
Sahu et al.~\cite{Sahu:2013jla} introduced a semi-empirical relationship based on the resonance phenomenon by considering the parent nucleus as a quantum two-body system of the ejected particle and the daughter nucleus. However, their calculated $\alpha$-decay half-lives are unsatisfactorily compared to the experimental results.
Based on the Sahu formula~\cite{Sahu:2013jla}, by considering a precise radius formula and an analytic expression for preformation factor, Zhang et al.~\cite{Zhang:2017pwm}
proposed an improved semi-empirical relationship for $\alpha$-decay half-lives (denoted as ZZCW). 
ZZCW is given as
\be
\log_{10} \left( T^{\rm ZZCW}_{1/2}[\text{s}] \right) = a \Za \Zd \sqrt{\mu_A/\Qa} + b \sqrt{\mu_A \Za \Zd} + c + d\, , \label{formula1}
\ee
where  $T_{\rm cal}[\text{s}]$ is the half-life measured in seconds 
and $\Za$ and $\Zd$ denote the proton number of alpha particle and daughter nucleus, respectively. 
The reduced mass number $\mu_A$ is defined by $\mu_A = \Aa \Ad/A_{\rm par}$, where $\Aa, \Ad$, and $A_{\rm par}$ are the mass number of the alpha particle, daughter nucleus, and parent nucleus, respectively.
For the detailed description of the parameters  $a$, $b$, $c$, and $d$ we refer to Zhang et al.~\cite{Zhang:2017pwm}.  They found that ZZCW significantly improved the accuracy of the $\alpha$-decay half-lives in comparison with experimental results. Based on the success, they predicted $\alpha$-decay half-lives of the $Z=118-121$ isotopes and a magic number effect at $N_{\rm par}=184$.

\item {UNIV}:
Single universal curve for cluster radioactivities and $\alpha$-decay was proposed by Poenaru et al.~\cite{Poenaru:2011zz} for even-even, even-odd, and odd-even parent nuclei. Due to the lack of successful experiment on cluster radioactivities of odd-odd parent nuclei, parameters for odd-odd parent nuclei were not provided. %~\cite{Poenaru:2011zz}.
The proposed formula for even-even isotopes (denoted as UNIV) is given as
\be
\log_{10} \left( T^{\rm UNIV}_{1/2}[\text{s}] \right)    = - \log_{10} P - \log_{10} S + c_{ee}, \label{formula5}
\ee
where $- \log_{10} P = 0.22873\sqrt{\mu_A \Zd \Za R_b} (\arccos \sqrt{r} - \sqrt{r(1-r)} )$, $\log_{10} S = -0.598(\Aa-1)$, and $c_{ee}=-22.16917$. $r=R_t/R_b$, $R_t=1.2249(\Ad^{1/3}+\Aa^{1/3})$, and $R_b=1.43998\Zd\Za/\Qa$.
This formula was obtained by plotting the sum of the decimal logarithm of the half-life and cluster preformation factor versus the decimal logarithm of the penetrability of external barrier~\cite{Poenaru:2011zz}. Akrawy and Poenaru~\cite{Akrawy:2017lpn} found that UNIV gives quite good results for the $\alpha$-decays of even-even nuclei and suggest that UNIV may be 
good for the $\alpha$-decay half-lives of superheavy nuclei.

\end{itemize}

\end{document}